\documentclass[a4paper,11pt]{article}
\usepackage{jheppub} % for details on the use of the package, please see the JINST-author-manual
\usepackage{lineno}
\usepackage{amsfonts}
\usepackage{soul}
\usepackage{xcolor}
\usepackage{subcaption}
\usepackage{orcidlink}

\newcommand{\nn}{\nonumber}

\arxivnumber{2412.08517}

\title{Smart Holes: Analogue black holes with the right temperature and entropy}

\author[a,b,c,d]{Jiayue Yang\orcidlink{0009-0004-0780-1465},}
\author[a,b,c]{Niayesh Afshordi\orcidlink{0000-0002-9940-7040},}
\author[b]{Mahdi Torabian\orcidlink{0000-0002-2993-913X},}
\author[e]{Seyed Akbar Jafari\orcidlink{0000-0003-3056-1063}} 
\author[b,f,g]{and G. Baskaran\orcidlink{0009-0009-5870-4442}}
\affiliation[a]{Department of Physics and Astronomy, University of Waterloo,\\ 
200 University Ave W, Waterloo, ON N2L 3G1, Canada}

\affiliation[b]{Perimeter Institute For Theoretical Physics, \\ 
31 Caroline St N, Waterloo, ON N2L 2Y5, Canada}
\affiliation[c]{Waterloo Centre for Astrophysics, University of Waterloo, \\
200 University Ave W, Waterloo, ON N2L 3G1, Canada}
\affiliation[d]{Department of Applied Mathematics, University of Waterloo,\\
200 University Ave W, Waterloo, ON N2L 3G1, Canada}
\affiliation[e]{II. Physikalisches Institut, RWTH Aachen University,\\ 52074 Aachen, Germany}
\affiliation[f]{The Institute of Mathematical Sciences, C.I.T. Campus,\\ Chennai 600 113, India}
\affiliation[g]{Department of Physics, Indian Institute of Technology,\\ Chennai 600 036, India}

\emailAdd{j43yang@uwaterloo.ca}
\emailAdd{nafshordi@pitp.ca}
\emailAdd{mahditorabian@g.harvard.edu}
\emailAdd{akbar.jafari@rwth-aachen.de}
\emailAdd{baskaran@imsc.res.in}

\abstract{In analogue gravity studies, the goal is to replicate black hole phenomena, such as Hawking radiation, within controlled laboratory settings. In the realm of condensed matter systems, this may happen in 2D tilted Dirac cone materials based on honeycomb lattice. In particular, we compute the entropy of this system, and find it has the same form as black hole Bekenstein-Hawking entropy, if an analogue horizon forms. Hence, these systems can be potential analogues of quantum black holes. We show that this entropy is primarily concentrated in the region where the tilt parameter is close to one, which corresponds to the location of the analogue black hole horizon. Additionally, when nonlinear effects are taken into account, the entropy is peaked in a small pocket of the Fermi sea that forms behind the analogue event horizon, which we call the \textit{Fermi puddle}.  We further refer to this new type of analogue black hole as a {\it smart hole}, since, in contrast to dumb holes, it can simulate both the correct temperature {\it and} entropy of general relativistic black holes.   These results provide an opportunity to illuminate various quantum facets of black hole physics in a laboratory setting.}


\begin{document}
\maketitle
\flushbottom

\section{Introduction}\label{sec1}

Making analogies between distinct systems is a very common and useful method in physics. In various approaches to quantum gravity, the concept of ``analogue gravity'' emerges as a tool to investigate aspects of gravitational theory in laboratories.  The origins of analogue gravity trace back to Unruh's groundbreaking paper in 1981 \cite{Unruh:1980cg}. In this work, he drew an analogy between the sonic horizon in trans-sonic fluid flow and the event horizon of gravitational black holes; events within the sonic horizon cannot be heard from outside. Unruh called these analogue black holes, ``dumb holes'', referring to the original meaning of the word ``dumb'', as unable to speak.  

Analogue gravity usually involves employing condensed matter systems, such as fluids, superfluids, and Bose-Einstein condensates, to replicate gravitational phenomena \cite{barcelo2011analogue,Visser:2001fe,Jacquet:2020bar,Fischer:2004bf,Fedichev:2003bv,Fedichev:2003id,Baak:2022hum}. As an intersection of condensed matter physics and gravitational physics \cite{Unruh:1980cg,barcelo2011analogue,Visser:2001fe,Jacquet:2020bar,Fischer:2004bf,Fedichev:2003bv,Fedichev:2003id,Baak:2022hum,Paulino:2022amx}, the study of analogue gravity has greatly advanced the development of condensed matter experiments and enhanced our comprehension of phenomena in curved spacetime. As the most mysterious objects in the universe, black holes are one of the most fascinating predictions of general relativity. To better understand the nature of gravity and manipulate artificial black holes in laboratories \cite{novello2002artificial},  black hole analogue models based on various systems have been put forward, such as sonic analogue black holes in dilute Bose-Einstein condensates \cite{PhysRevLett.105.240401,Garay:1999sk,Garay:2000jj},  analogue black holes in dipolar condensates \cite{Ribeiro:2022gln},  analogue black holes in finite quasi-one-dimensional Bose-Einstein condensates \cite{Ribeiro:2021fpk}, optical analogue black holes \cite{Unruh:2003ss,Rosenberg:2020jde,PhysRevA.86.063821,Gaona-Reyes:2017mks},  analogue black holes in flowing dielectrics \cite{Novello:2002ed},   analogue black holes in superfluid \cite{Volovik:2000ua,Volovik:2002ci},  and analogue black holes in fermionic systems \cite{Huang:2017eio,Konye:2022wds,Kang:2019pch,sims2023analogous}. Quantum matter systems based on the Sachdev–Ye–Kitaev (SYK) model \cite{Sachdev:1992fk,Kitaev} also have emerged as a promising framework for mimicking black holes \cite{Franz:2018cqi} in the context of holographic duality \cite{Susskind:1994vu,Maldacena:1997re,Witten:1998qj}. 

A famous consequence of the quantum field theory in curved space-time is Hawking radiation \cite{Hawking:1974rv,Hawking:1975vcx},  the notion that black holes are not completely black, but could emit radiation, through quantum processes, characterized by the Hawking temperature. However, detecting Hawking radiation of astrophysical black holes in the universe is extremely challenging (if not impossible), so diverse analogues of Hawking radiation were proposed \cite{Huhtala:2001cj,volovik2016black,volovik2017lifshitz,Weinfurtner:2010nu,Steinhauer:2014dra,MunozdeNova:2018fxv,Kolobov:2019qfs,Euve:2021mnj,Bagchi:2023vzx,Robertson:2012ku,Drori:2018ivu,Dardashti:2015tgp,Stone:2012cx,Subramanyan:2020fmx,Ma:2024ivd,Horner:2022sei,Yang:2019kbb,Shi:2021nkx,Sabsovich:2021khz,Maertens:2022yvk,Steinhauer:2015saa,Tian:2020bze}. For the typical analogue black hole (``dumb hole" according to Unruh), the entropy does not match the widely accepted horizon entropy of a quantum black hole, as predicted by Bekenstein \cite{Bekenstein:1972tm,Bekenstein} and Hawking \cite{Hawking:1975vcx}, making them incomplete analogues \cite{Unruh:1980cg}. However, some analogue black holes may exhibit the correct entropy, and we refer to these as ``smart holes" (contrasting with the more commonly used meaning of the word ``dumb''). A smart hole is thus an analogue black hole that possesses both the correct temperature and entropy, making it a more accurate representation of a quantum black hole.

A very fascinating and promising analogue gravity system can be inspired by  graphene's honeycomb lattice, a hexagonal arranged two-dimensional single layer of carbon atoms with Dirac cones at the six vertices of the hexagonal Brillouin zone \cite{CastroNeto:2007fxn,barcelo2011analogue,Iorio:2011yz,Iorio:2013ifa,Iorio:2014pwa,Acquaviva:2022yiq}, as shown in Figure \ref{graphene}. Graphene is regarded as a zero-bandgap semiconductor. The Dirac cone points, where the valence band touches the conduction band, are an important feature of the electronic band structure of graphene, as shown in Figure \ref{full band}. The low-energy excitations around the Dirac points can be regarded as massless Dirac fields governed by the Dirac equation in 2+1 dimensions \cite{Vozmediano:2008zz,Kosinski:2012vt}.  Recent studies indicate that in the case of anisotropic hoppings \cite{yekta2023tunning}, external electromagnetic fields \cite{goerbig2011electronic}, mechanical deformation \cite{goerbig2008tilted} and chemical strain \cite{geilhufe2018chemical}, the originally upright Dirac cones may undergo tilting. Various systems, for example quinoid-type graphene \cite{goerbig2008tilted}, partially hydrogenated graphene \cite{lu2016tilted}, 8Pmmn Borophene \cite{yekta2023tunning}, $1{T}^{\ensuremath{'}}\text{\ensuremath{-}}\mathrm{Mo}{\mathrm{S}}_{2}$ \cite{tan2021anisotropic}, and organic conductor \cite{hirata2016observation,suzumura2016analysis}, can exhibit tilted Dirac cones with substantial tilts \cite{zheng2020tilted}. In materials with tilted Dirac cones, the geometry of the zero-energy Fermi surface is expected to change with variations in the tilting parameter \cite{Huang:2017eio,Soluyanov:2015mvf,Milievi2018TypeIIIAT,fan2023two}. See Figure \ref{linear dispersion relation} for different types of tilted Dirac cones, the types are categorized by the shape of the zero-energy Fermi surface.

The symmetry argument behind the tilting of the Dirac cone in two space dimensions is as follows: the low-energy Dirac theory in pristine graphene is rotationally symmetric in two-dimensions (the $xy$ plane). Now imagine breaking the rotational symmetry by introducing a preferred direction in this plane. In this case, the only allowed rotation around $z$-axis is rotation by $\pi$ which in the solid state jargon is denoted by a two-fold rotation axis and is called $C_2$.  Additionally, there is a reflection in a plane that contains this two-fold axis and is vertical to the two-dimensional space of the system.  If both directions along the preferred axis of the $xy$ plane are identical, then there will be another reflection plane again vertical to the $xy$ plane that contains the $C_2$ symmetry axis. This point group symmetry is called a $C_{2v}$ symmetry and  can be achieved by introducing a term of the form $v_F\boldsymbol{p}.\boldsymbol{\zeta}\sigma_0$ in the Bloch Hamiltonian where $\sigma_0$ is the unit $2\times2$ matrix, $\boldsymbol{p}=(p_x,p_y)$ is the two-dimensional momentum, $\boldsymbol{\zeta}=(\zeta_x,\zeta_y)$ is the dimensionless parameter that quantifies the amount of tilting and $v_F$ is the natural velocity scale, the Fermi velocity~\cite{yekta2023tunning}. This logic is quite powerful and can be used to design other platform hosting tilted cones: A simple honeycomb lattice of (bosonic) LC resonators on the honeycomb lattice features a ``microwave cone" that can be tilted by breaking the symmetry $C_{6v}$ of the lattice down to $C_{2v}$ that can be achieved by introducing additional second neighbour couplings in one preferred directions~\cite{Motavassal2021}.  The advantage of fermion systems is that the excitations in fermionic systems will be electrons and holes that are charged and interact via Coulomb force. Therefore one can study the interplay between the Coulomb interaction and the new spacetime that arises from tilting of the Dirac cones (potentially mimicking gravitational interactions). 

Dirac cones are not limited to honeycomb-based graphene-like systems. They can also arise from spin-orbit interactions. For such systems in two space dimensions, the Dirac matrices representing the Clifford algebra will be the Pauli matrices associated with the actual spin (as opposed to the sublattice or pseudo-spin degree of freedom). Examples of such systems are the helical Dirac cones on the surface of topological insulators. It has been shown that an in-plane magnetic influence in such systems, directly couples to the Dirac cone in the ``tilt channel", and becomes a knob to adjust the tilt of the Dirac cone by tens of percents~\cite{Jafari2023Moving} which is quite comparable to the $\sim 40\%$ tilt in pristine 8Pmmn-borophene. Substituting certain boron atoms in the above material with carbon can increase the tilting up to $\sim 60\%$ \cite{yekta2023tunning}. But the spin-orbit coupled Dirac cones have the advantage that a comparably large tilting can be produced and controlled with external magnetic influence~\cite{Jafari2023Moving}. 

Based on the work of Volovik \cite{nissinen2017type,volovik2016black, volovik2018exotic,volovik2017lifshitz,Huhtala:2001cj}, the emergent spacetime metric of tilted Dirac cone is given by the so called Painlevé-Gullstrand (PG) line element \cite{Jafari:2019ufn,Jalali-Mola:2019njn,farajollahpour2019solid,Moradpouri:2022mhi}
\begin{align}\label{metric}
    ds^2=-v_F^2dt^2+(d\boldsymbol{r}-\boldsymbol{\zeta}v_Fdt)^2,
\end{align}
where $v_F$ is the Fermi velocity   playing the role of speed of light $c$ in emergent spacetime and $\boldsymbol{\zeta}$ is the dimensionless parameter controlling the tilt. Recently, it has been shown that the above tilt parameter $\boldsymbol{\zeta}$ in spin-orbit coupled Dirac cones can be tuned by in-plane magnetization/magnetic field~\cite{Jafari2023Moving}.  A simple argument obtaining the above metric from the term $v_F\boldsymbol{p}\cdot\boldsymbol{\zeta}\sigma_0$ in the Bloch Hamiltonian is given in Ref.~\cite{Jafari2023Moving}. In this setup, crossing $|\boldsymbol{\zeta}|=1$ would correspond to crossing the event horizon of the analogue black hole. 

This paper is organized as follows. In section \ref{sec band}, we review the band structure of tilted Dirac cone materials and show the first and second order expansion of the dispersion near the Dirac points. In section \ref{sec linear}, we consider the linear dispersion to compute the entropy of analogue black hole constructed by tilted Dirac cone materials. In section  \ref{sec num}, we present the numerical results for both the entropy density and total entropy. We find the  temperature and total entropy of tilted Dirac cone material, under certain assumptions, have the same form as a 2+1 dimensional black hole. We summarize our results in section \ref{sec con}.

\section{Band structure of tilted Dirac cone materials}\label{sec band}
Building on the emergent metric structure introduced in section \ref{sec1}, where $\boldsymbol{\zeta}$ governs the tilt of the Dirac cone and gives rise to an effective curved spacetime, we can obtain the corresponding Hamiltonian for the tilted Dirac cone system.  To better understand the microscopic origin of this emergent geometry, we now turn to the condensed matter perspective. In this section, we  review the Hamiltonian and the band structure of tilted Dirac cone materials.
Following the work  \cite{yekta2023tunning}, the basic idea is that the nearest-neighbour hopping described by the tight-binding model is responsible for the formation of upright Dirac cones. When the anisotropic next-nearest-neighbor hopping is introduced, the Dirac cones tilt along one direction, resulting in an anisotropic band structure near the Dirac points.

\begin{figure}[!h]
    \centering
\includegraphics[width=0.7\textwidth]{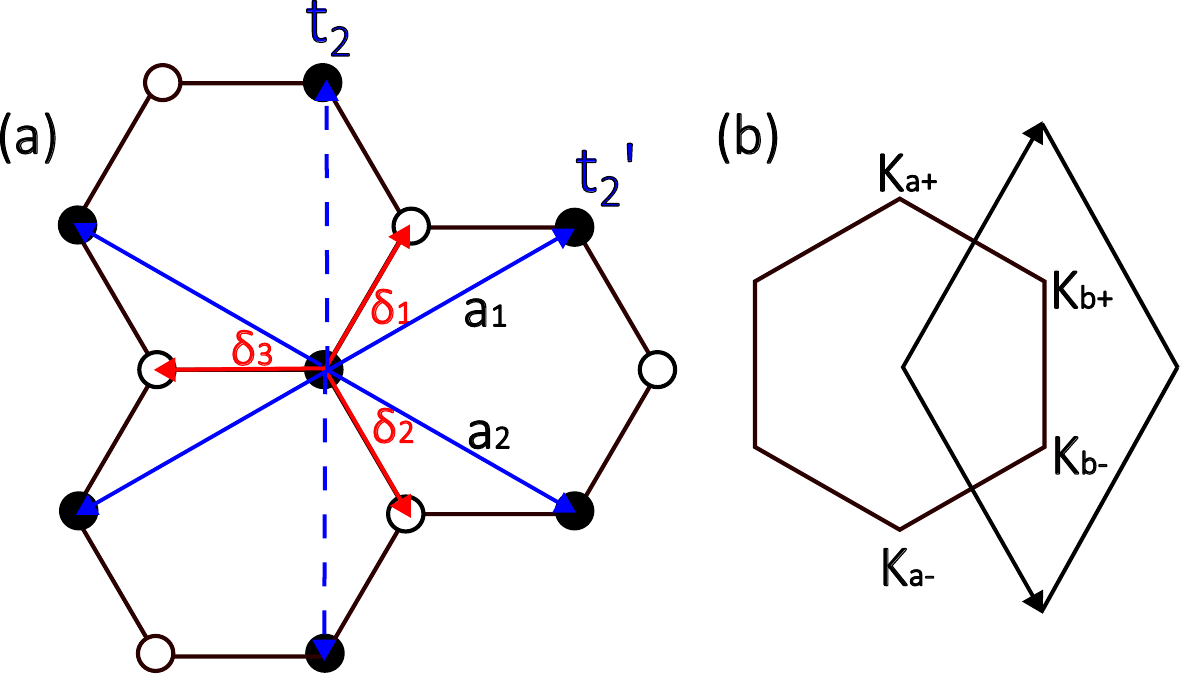}
    \caption{(a) From honeycomb lattice to honeycomb graph: The honeycomb lattice with two sublattices (black and white) and lattice constant $a$. The primitive lattice vectors are  \( \mathbf{a}_1 = a \left( \frac{\sqrt{3}}{2}, \frac{1}{2} \right) \) and \( \mathbf{a}_2 = a \left(  \frac{\sqrt{3}}{2}, -\frac{1}{2} \right) \).
Three red vectors connecting to nearest neighbours are denoted by $\boldsymbol{\delta_i} 
(i = 1, 2, 3)$. Hopping along nearest neighbors is materialized in graphene. The lattice will be promoted to a graph that is effectively realized in 8Pmmn borophene by adding blue links connecting second nearest neighbours.  Two dashed blue vectors correspond to hopping parameter $t_2$ and four solid blue vectors have hopping parameter $t_2'$. The tilting arises from the {\it difference} between $t_2$ and $t'_2$. (b)  First Brillouin
zone in  reciprocal space with  K-points.}
    \label{graphene}
\end{figure}

The honeycomb lattice structure of graphene (or 8Pmmn borophene) in real space and reciprocal space are shown in Figure \ref{graphene}. With a 2D hexagonal lattice structure, graphene possesses two carbon atoms within each unit cell. There are  two sublattices called sublattice A and sublattice B respectively  in graphene. The lattice constant of graphene is denoted by \(a\), and the primitive lattice vectors can be expressed as \( \mathbf{a}_1 = a \left( \frac{\sqrt{3}}{2}, \frac{1}{2} \right) \) and \( \mathbf{a}_2 = a \left(  \frac{\sqrt{3}}{2}, -\frac{1}{2} \right) \) in real space  \cite{baskaran2012physics}. 
An atom in sublattice A is connected to the nearest-neighbour atoms in sublattice B via three vectors $\boldsymbol{\delta}_i$ given as
\begin{align}
\boldsymbol{\delta}_1=\frac{a}{2\sqrt{3}}(1, \sqrt{3}), \quad \boldsymbol{\delta}_2=\frac{a}{2\sqrt{3}}(1,-\sqrt{3}), \quad \boldsymbol{\delta}_3=-\frac{a}{\sqrt{3}}(1,0).
\end{align}

In the tight-binding model, we just consider the nearest-neighbour hopping, which is a hopping process from a site in one sublattice to one in the other sublattice. 
So we have the Hamiltonian in real space
\begin{align}
    \hat{H}=-t \sum_{\langle i j\rangle, \sigma}\hat{a}_{i, \sigma}^{\dagger} \hat{b}_{j, \sigma}+h.c. ,
\end{align}
 where  $t$ represents the hopping amplitude 
 or the hopping integral, which we suppose is a real number, $a_{i,\sigma}^{\dagger}$ creates a particle with spin state $\sigma$ at site $i$, and 
 $b_{j,\sigma}$ annihilates a particle with spin state $\sigma$ at site $j$.
 %, and $h.c.$ denotes the Hermitian conjugate term. 
The Hamiltonian in momentum space, after Fourier transformation, is
\begin{align}
H_0(\boldsymbol{k})=\left(\begin{array}{cc}
0 & f(\boldsymbol{k}) \\
f^*(\boldsymbol{k}) & 0
\end{array}\right),
\end{align}
where $\boldsymbol{k} $ is the  wave vector, and the form factor $f(\boldsymbol{k})$, responsible for the formation of the upright Dirac Cones, is given by
\begin{align}
  & f(\boldsymbol{k})=\sum_{\boldsymbol{\delta}}t e^{i \boldsymbol{k} \cdot \boldsymbol{\delta}} =t\left(e^{i\left(k_x \frac{a}{2 \sqrt{3}}+k_y \frac{a}{2}\right)}+e^{i\left(k_x \frac{a}{2 \sqrt{3}}-k_y \frac{a}{2}\right)}+e^{-i\left(k_x  \frac{a}{\sqrt{3}}\right)}\right). 
\end{align}
The band structure of normal graphene with upright Dirac cones is given by
\begin{align}
 E_0=\pm | f(\boldsymbol{k})| = \pm t\sqrt{3 + 4\cos\left(\frac{\sqrt{3}k_xa}{2}\right) \cos\left(\frac{k_ya}{2}\right) + 2\cos(k_ya)}, 
\end{align}
where $t$ is the hopping parameter and $a$ is the lattice constant, the $\pm$ means the upper (+) and lower ($-$) branches. 

\begin{figure}[!h]
    \centering
\includegraphics[width=0.7\textwidth]{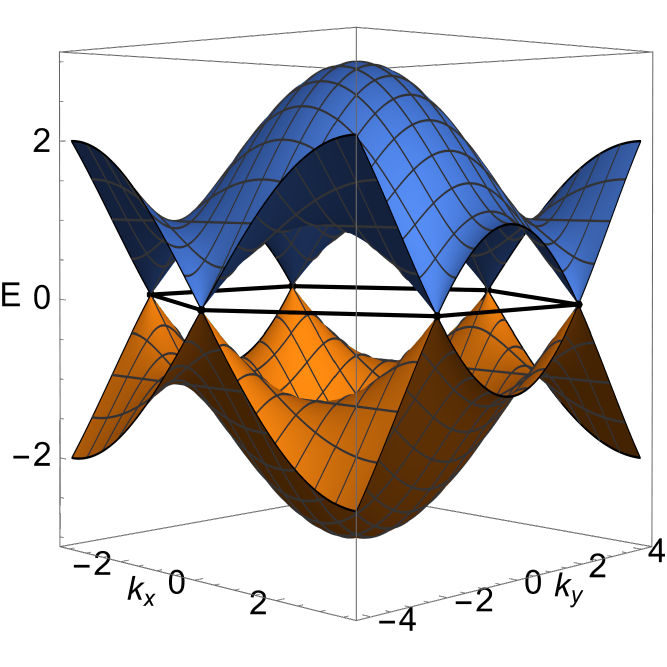}
    \caption{Electronic band structure of graphene.  Six Dirac points, or $K$ points (vertices of the hexagonal first Brillouin zone) are shown in the figure.  }
    \label{full band}
\end{figure}

What makes these Dirac cones particularly interesting is that they can be tilted, analogous to the tilting of light cones in the context of curved spacetime. This tilting can arise from the anisotropy of hopping. For instance, consider the second nearest-neighbour hopping with two distinct hopping parameters denoted by $t_2'$ and $t_2$, as shown in Figure \ref{graphene}. Since the second nearest-neighbour hoppings connect sites on the same sublattice, the corresponding form factor $f_{\text {tilt}}(\boldsymbol{k})=f_{t_2'}(\boldsymbol{k})+f_{ {t_2 }}(\boldsymbol{k})$ is the diagonal term of Hamiltonian, which is given by \cite{yekta2023tunning}
 \begin{align}
     f_{t_2'}(\boldsymbol{k})
&= t_2'\left(e^{ia\left(\frac{{k_x\sqrt{3}}}{2} + \frac{{k_y}}{2}\right)} + e^{ia\left(\frac{{k_x\sqrt{3}}}{2} - \frac{{k_y}}{2}\right)} + e^{ia\left(-\frac{{k_x\sqrt{3}}}{2} + \frac{{k_y}}{2}\right)}+ e^{ia\left(-\frac{{k_x\sqrt{3}}}{2} - \frac{{k_y}}{2}\right)} \right),   \\
  f_{t_2}(\boldsymbol{k})&=t_2\left( e^{ia k_y} + e^{-ia k_y}\right),\\
  f_{\text {tilt}}(\boldsymbol{k})&= 4t_2'\cos\left(\frac{{\sqrt{3} k_xa}}{2}\right) \cos\left(\frac{{k_ya}}{2}\right) +2 t_2\cos(k_ya).
 \end{align}
 Then, the Hamiltonian for tilted Dirac cone materials is given by
\begin{align}\label{ham2}
    H_{\text {eff }}(\boldsymbol{k})=\left(\begin{array}{cc}
f_{\text {tilt}}(\boldsymbol{k}) & f(\boldsymbol{k}) \\
f^*(\boldsymbol{k}) & f_{\text {tilt}}(\boldsymbol{k})
\end{array}\right),
\end{align}
from which we can obtain the full electronic band structure of tilted Dirac cone material as follows \begin{align}\label{full}
E_{\pm}=& f_{\text {tilt}}(\boldsymbol{k})\pm | f(\boldsymbol{k})|\nn\\
=&4t_2' \cos\left(\frac{\sqrt{3}k_xa}{2}\right) \cos\left(\frac{k_ya}{2}\right) + 2t_2 \cos(k_ya)\nn\\
\pm& t \sqrt{3 + 4\cos\left(\frac{\sqrt{3}k_xa}{2}\right) \cos\left(\frac{k_ya}{2}\right) + 2\cos(k_ya)},
\end{align}
where $k_x$ and $k_y$ have period $\frac{4 \pi}{\sqrt{3}a}$ and $\frac{4 \pi}{a}$ respectively. The band structure without tilt ({\it i.e.} upright cone with $t_2 = t_2' = 0$) is depicted in Figure \ref{full band}. As can be seen, there are six Dirac points, also known as $K$ points.
We expand the form factors around the $K$ points $K_a^{ \pm}=(0,\pm \frac{4\pi}{3a})$ to obtain the Dirac cone dispersion relation.

After expanding $f \left(K_a^{ \pm}+\boldsymbol{k}\right)$ to the second order, we get
\begin{align}
  f\left(K_a^{ \pm}+\boldsymbol{k}\right)
&=f\left(K_a^{\pm}\right)+\left.\frac{\partial f }{\partial k_x}\right|_{k=K_a^{ \pm}} k_x+\left.\frac{\partial f}{\partial k_y}\right|_{k=K_a^{ \pm}} k_y\nn\\
 &+\frac{1}{2}\left.\frac{\partial^2 f }{\partial k_x^2}\right|_{k=K_a^{ \pm}} k_x^2+\frac{1}{2}\left.\frac{\partial^2 f }{\partial k_y^2}\right|_{k=K_a^{ \pm}} k_y^2+\left.\frac{\partial^2 f }{\partial k_x\partial k_y}\right|_{k=K_a^{ \pm}} k_xk_y\nn\\&=t\left(-\frac{\sqrt{3}a
}{2}  i\left(k_x \mp i k_y\right)+\frac{1}{8} a^2(-k_x^2+k_y^2)\mp\frac{1}{4} i a^2 k_x k_y\right).
\end{align}
If we keep it to first order, we find
\begin{align}\label{f1}
  f^{(1)}\left(K_a^{ \pm}+\boldsymbol{k}\right)=-i\frac{\sqrt{3}at
}{2}  \left(k_x \mp i k_y\right).
\end{align}
 
The overall factor of $-i=e^{-i \pi / 2}$ can be absorbed through rotating the overall phase of the wave function in one sublattice. Then, the Hamiltonian around the Dirac points is computed as
\begin{align}
    H_{\text{upright}}(\boldsymbol{k})=\hbar v_F \begin{bmatrix}
0 & k_x \mp i k_y \\
k_x \pm i k_y & 0
\end{bmatrix},
\end{align}
where $v_F=\frac{\sqrt{3}}{2}at$ is the Fermi velocity of electrons at the Dirac Cone.

For the case of tilted Dirac cone, after expanding $f_{\text {tilt}} \left(K_a^{ \pm}+\boldsymbol{k}\right)$ to second order, we find
\begin{align}\label{ftilt}
f_{\text {tilt}} \left(K_a^{ \pm}+\boldsymbol{k}\right)
    &=f_{\text {tilt}}\left(K_a^{+}\right)+\left.\frac{\partial f_{\text {tilt}} }{\partial k_x}\right|_{k=K_a^{ \pm}} k_x+\left.\frac{\partial f_{\text {tilt}} }{\partial k_y}\right|_{k=K_a^{ \pm}} k_y\nonumber\\
    &+\frac{1}{2}\left.\frac{\partial^2 f_{\text {tilt}} }{\partial k_x^2}\right|_{k=K_a^{ \pm}} k_x^2+\frac{1}{2}\left.\frac{\partial^2 f_{\text {tilt}} }{\partial k_y^2}\right|_{k=K_a^{ \pm}} k_y^2+\left.\frac{\partial^2 f_{\text {tilt}} }{\partial k_x\partial k_y}\right|_{k=K_a^{ \pm}} k_xk_y\nonumber\\
    &=(-2 t_2' - t_2)\mp \sqrt{3} a( t_2' -  t_2)k_y+ \frac{3 t_2'}{4}  a^2 k_x^2+\frac{1}{4}  (t_2' + 2 t_2) a^2 k_y^2.
\end{align}
Keeping the first order expansion, we find
\begin{align}\label{ftilt1}
f_{\text {tilt}}^{(1)} \left(K_a^{ \pm}+\boldsymbol{k}\right)
     &=(-2 t_2' - t_2)\mp \sqrt{3} a( t_2' -  t_2)k_y. 
\end{align}
Plugging Equations (\ref{f1}) and (\ref{ftilt1}) into Equation (\ref{ham2}), the free Hamiltonian of tilted Dirac cone materials can then be expressed as  \cite{Moradpouri:2022mhi}
\begin{align}\label{ham1}
    H= v_F
    \begin{bmatrix}
    \boldsymbol{p}\cdot \boldsymbol{\zeta}&p_x\mp ip_y\\
    p_x\pm ip_y&  \boldsymbol{p}\cdot \boldsymbol{\zeta}
    \end{bmatrix},
\end{align}
where $  \boldsymbol{p}$ is the momentum vector and $ \boldsymbol{\zeta}$ is the tilt parameter controlling the tilt of the Dirac cone.
We compute the tilt parameter $ \boldsymbol{\zeta}$ by comparing equations (\ref{ham1}) and (\ref{ham2}) with the first order approximation $f^{(1)}$ in (\ref{f1}) and $f_{\text {tilt}}^{(1)} $ in (\ref{ftilt1}) \cite{yekta2023tunning}:
\begin{align}
    \zeta_x=0,\qquad \zeta_y=\mp 2\frac{ t_2' -  t_2}{t}.
\end{align}
If we define the energy at the point of contact of the Dirac cones as the reference plane of zero energy, i.e., let $(-2 t_2' - t_2)=0$ in equation (\ref{ftilt1}), we get $t_2=-2t_2'$.  Then, the tilt parameter becomes
$\zeta_y= \mp 6t_2'/t $.

\begin{figure}[ht!]
    \centering
    \begin{subfigure}[b]{0.22\textwidth}
        \centering
        \includegraphics[width=\textwidth]{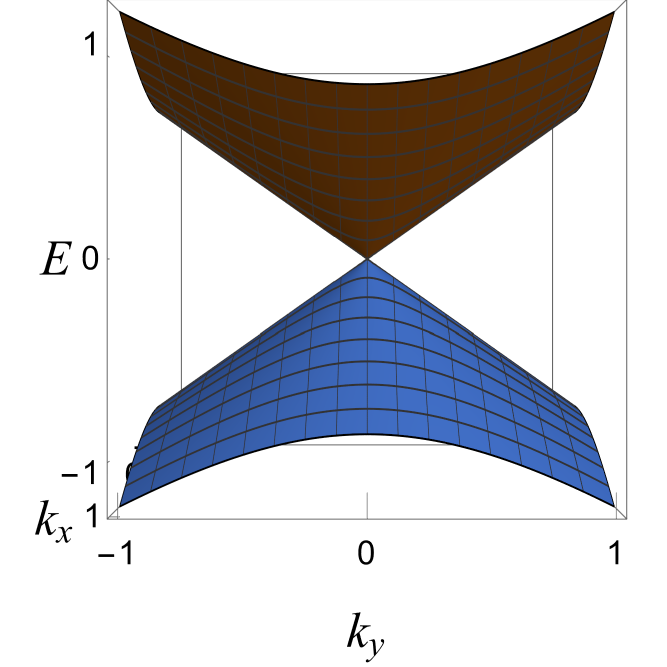}
     
    \end{subfigure}
    \hfill
    \begin{subfigure}[b]{0.24\textwidth}
        \centering
        \includegraphics[width=\textwidth]{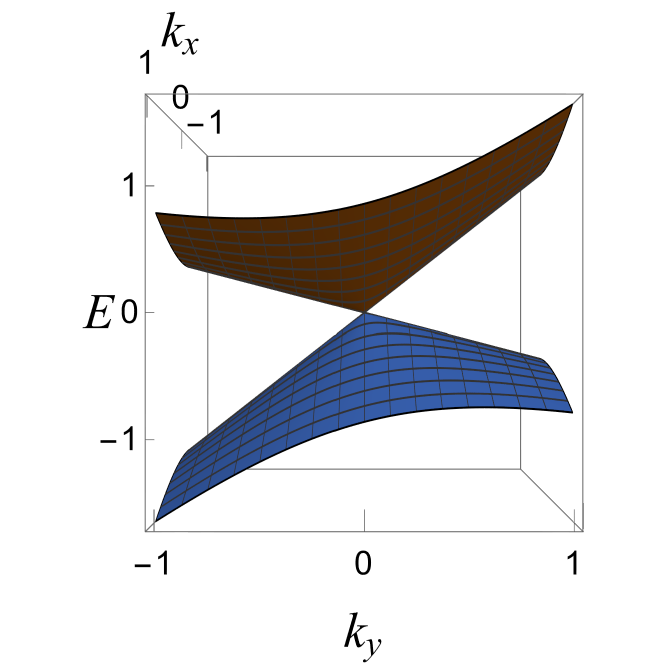}
      
    \end{subfigure}
    \hfill
    \begin{subfigure}[b]{0.24\textwidth}
        \centering
        \includegraphics[width=\textwidth]{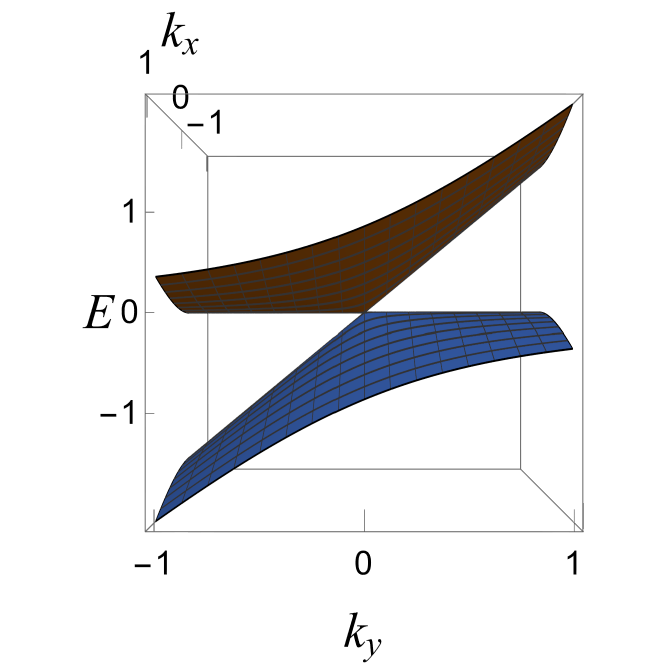}
      
    \end{subfigure}
      \hfill
    \begin{subfigure}[b]{0.24\textwidth}
        \centering
        \includegraphics[width=\textwidth]{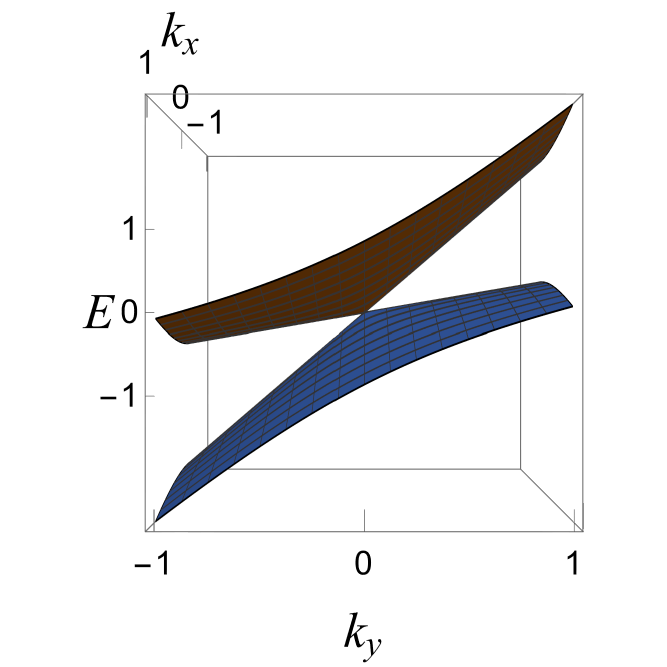}
       
    \end{subfigure}
    
     \medskip
     
          \begin{subfigure}[b]{0.22\textwidth}
        \centering
        \includegraphics[width=\textwidth]{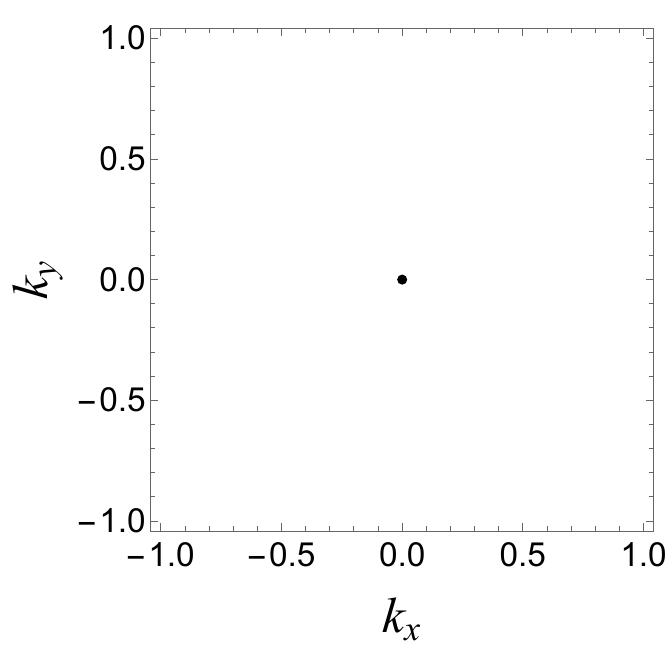}
        \caption{Tilt  $\zeta=0$}
       
    \end{subfigure}
    \hfill
    \begin{subfigure}[b]{0.22\textwidth}
        \centering
        \includegraphics[width=\textwidth]{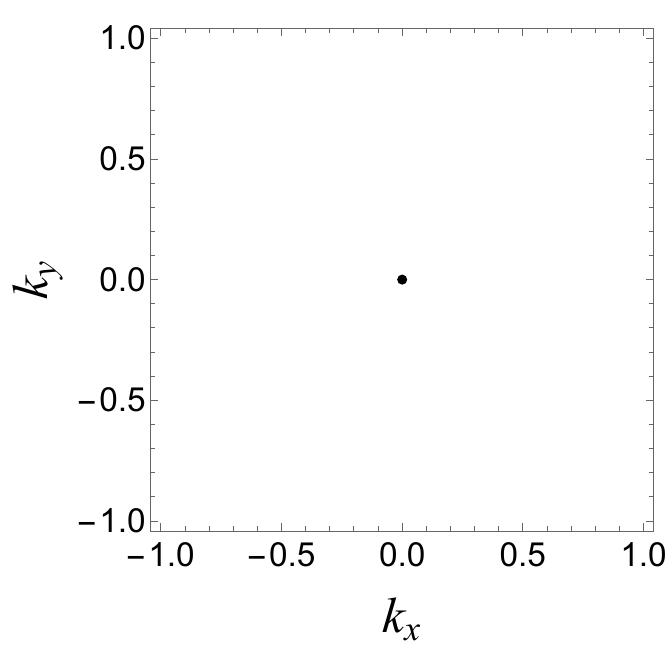}
        \caption{Tilt $\zeta=1/2$}
    \end{subfigure}
    \hfill
    \begin{subfigure}[b]{0.22\textwidth}
        \centering
        \includegraphics[width=\textwidth]{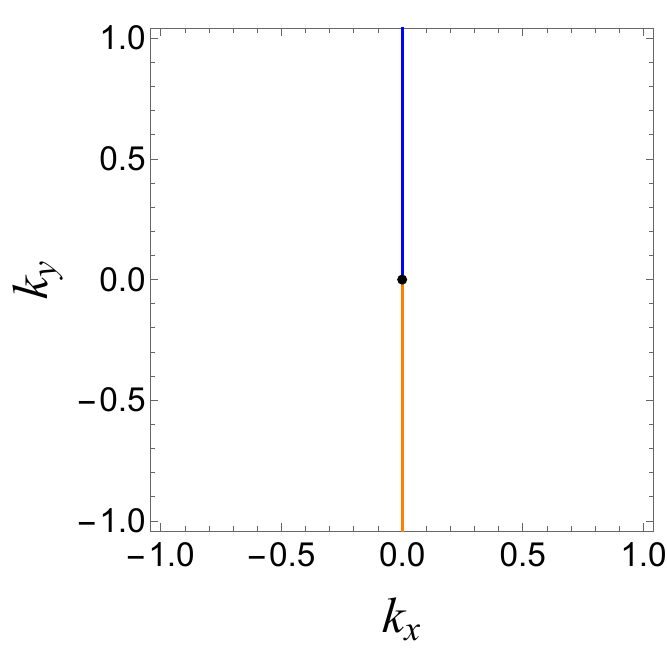}
        \caption{Tilt  $\zeta=1$}
    \end{subfigure}
      \hfill
    \begin{subfigure}[b]{0.22\textwidth}
        \centering
        \includegraphics[width=\textwidth]{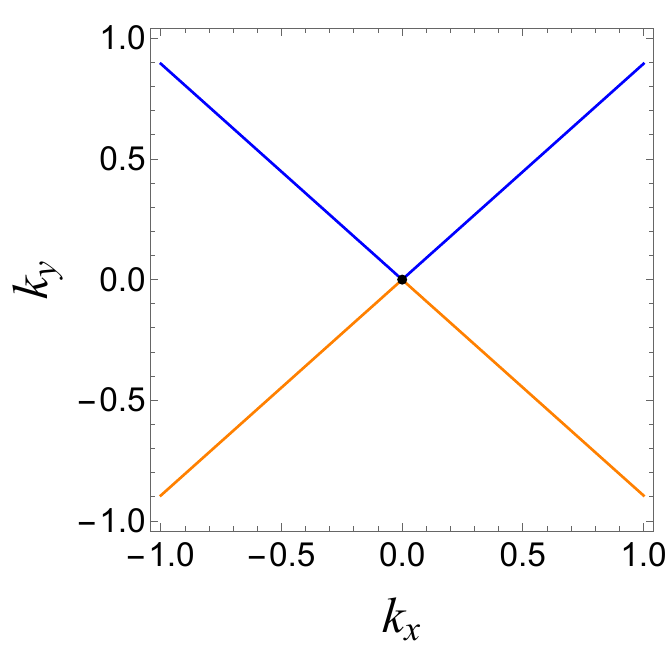}
        \caption{Tilt  $\zeta=3/2$}
    \end{subfigure}
    \caption{ The linear dispersion relation of the tilted Dirac cones. Top: dispersion relations with orange upper branch and blue lower branch. Bottom: zero-energy Fermi surface. (a) Type-I upright Dirac cone with 
 $ \zeta=0$,  (b)  Type-I tilted Dirac cone with 
 $ \zeta=1/2$, and (c) Type-III critically tilted Dirac cone with
 $ \zeta=1$ (d)  Type-II over-tilted  Dirac cone with $ \zeta=3/2$. } 
    \label{linear dispersion relation}
\end{figure}

The linear dispersion relation of the tilted Dirac cone is computed as 
\begin{align}\label{linearE}
    E=sv_Fp+v_F \boldsymbol{p}\cdot\boldsymbol{\zeta}=sv_Fp+v_Fp\zeta_y \sin{\theta}=s t\frac{\sqrt{3}}{2}ak+t_2'( \mp 3 \sqrt{3}\sin(\theta) ak,
\end{align}
where $p\equiv|\boldsymbol{p}|$, $k\equiv|\boldsymbol{k}|$, $\zeta\equiv |\boldsymbol{\zeta}|=\zeta_y$, $\theta$ is defined as the polar angle of $\boldsymbol{p}$ (the angle between $\boldsymbol{p}$ and $p_x$ component), and   $s=\pm 1$ indicates upper ``+''and lower ``$-$'' branches of the energy band structure.  Different types of the Dirac cone, which can be categorized based on the geometry of their Fermi surface \cite{Huang:2017eio,Soluyanov:2015mvf,Milievi2018TypeIIIAT,fan2023two}, are shown in Figure \ref{linear dispersion relation} (note that we set $t=1$, $a=1$ and $\hbar=1$ in plotting the band structure).  For  $\zeta=0$, we recover the standard (upright) Dirac cone. For  $0<\zeta<1$, the  Dirac cone becomes tilted. These two cases, whose zero-energy Fermi surface is a single point, are called type-I Dirac cone. When the Dirac cone is critically tilted with $\zeta=1$, the zero-energy Fermi surface becomes a straight line, known as the Dirac line. This is called type-III  Dirac cone. When $\zeta>1$, the zero-energy Fermi surface becomes two intersecting straight lines, and such an over-tilted Dirac cone is known as type-II Dirac cone.

We can use the condition $t_2=-2t_2'$  to further simplify equation (\ref{ftilt})  and get the generalized quadratic dispersion relation
\begin{align}
f_{\text {tilt}}\left(K_a^{ \pm}+\boldsymbol{k}\right)
    &=\mp 3\sqrt{3} a  t_2'k_y+ \frac{3 }{4}t_2'  a^2 k_x^2-\frac{3}{4}  t_2'  a^2 k_y^2.
\end{align}
Then, the dispersion relation to second order in polar coordinates is 
\begin{align}\label{quadratic}
    E& =f_{\text {tilt}}(\boldsymbol{k})+s | f(\boldsymbol{k})|\nonumber\\
    &=t_2'\Big( \mp 3 \sqrt{3}\sin(\theta) ak +\frac{3}{4} \cos(2 \theta)a^2k^2 \Big)+s t\Big(\frac{\sqrt{3}}{2}ak\pm\frac{\sin{(3\theta)}}{8} a^2k^2\Big).
\end{align}

\begin{figure}[ht!]
    \centering
    \begin{subfigure}[b]{0.24\textwidth}
        \centering
        \includegraphics[width=\textwidth]{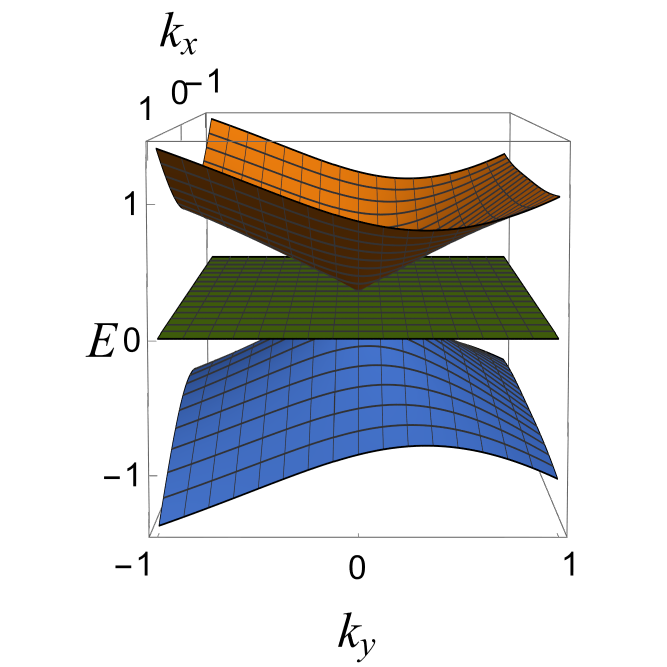}
        \label{fig:image1}
    \end{subfigure}
    \hfill
    \begin{subfigure}[b]{0.24\textwidth}
        \centering
        \includegraphics[width=\textwidth]{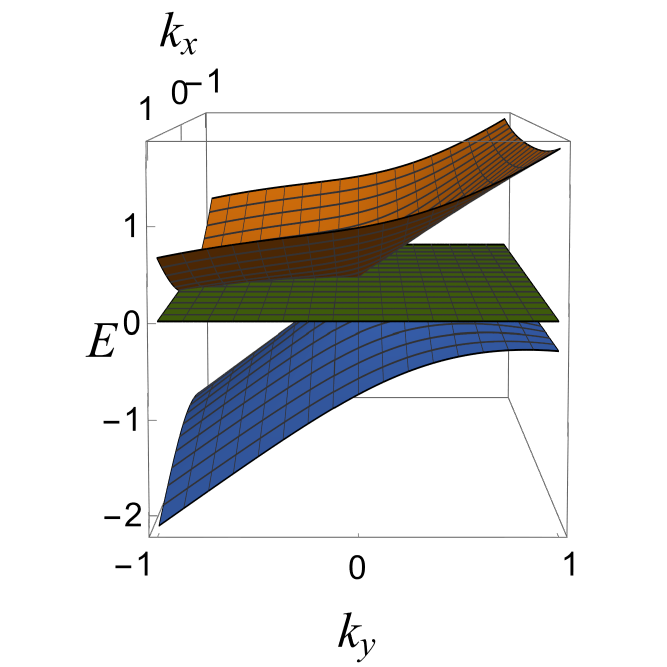}
        \label{fig:image2}
    \end{subfigure}
    \hfill
    \begin{subfigure}[b]{0.24\textwidth}
        \centering
        \includegraphics[width=\textwidth]{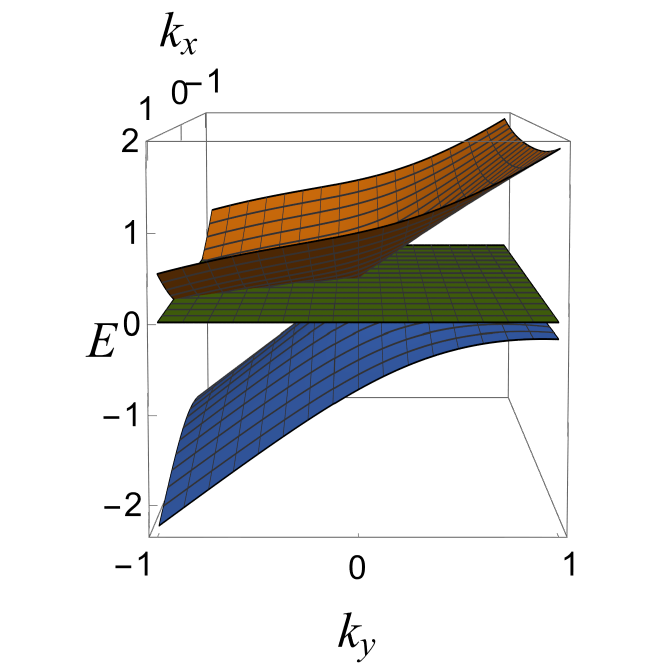}
        \label{fig:image3}
    \end{subfigure}
      \hfill
    \begin{subfigure}[b]{0.24\textwidth}
        \centering
        \includegraphics[width=\textwidth]{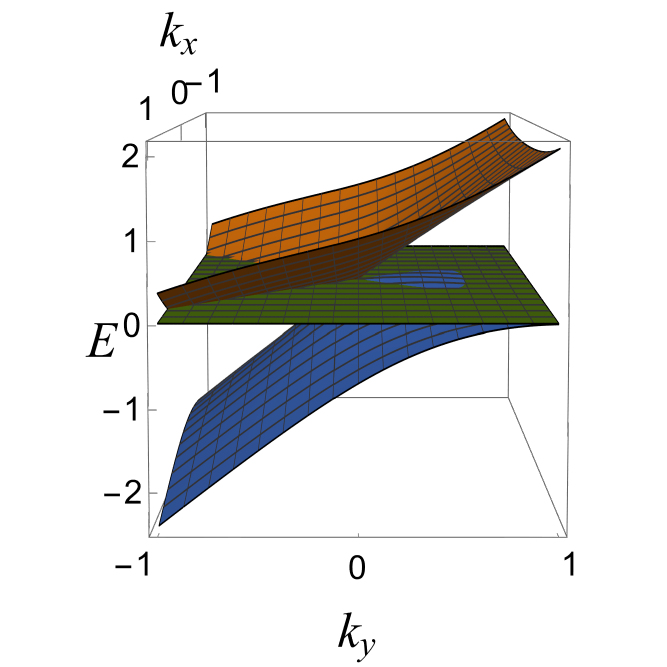}
        \label{fig:image3}
    \end{subfigure}
    
     \medskip
     
          \begin{subfigure}[b]{0.22\textwidth}
        \centering
        \includegraphics[width=\textwidth]{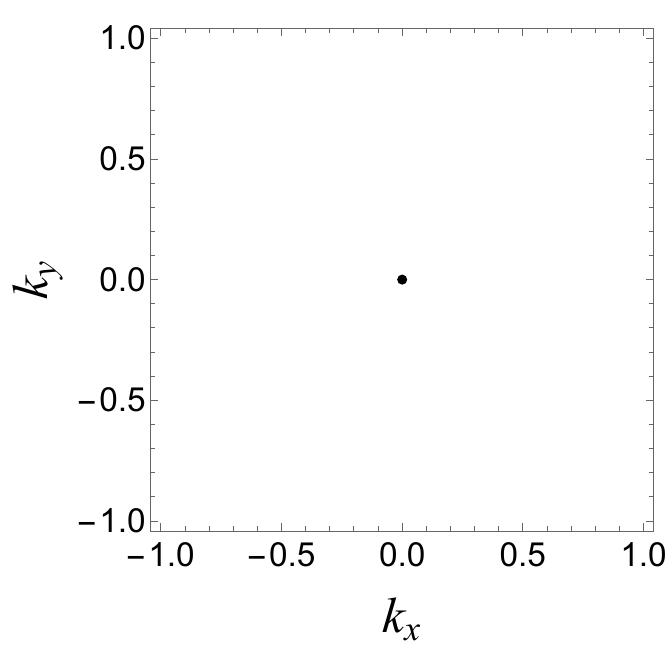}
        \caption{Tilt  $\zeta=0$}
        \label{fig:image1}
    \end{subfigure}
    \hfill
    \begin{subfigure}[b]{0.22\textwidth}
        \centering
        \includegraphics[width=\textwidth]{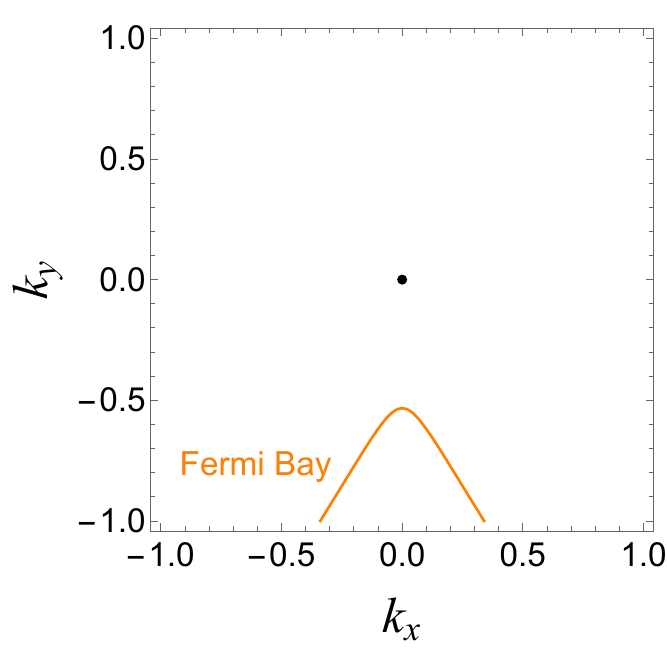}
        \caption{Tilt $\zeta=6/7$}
        \label{fig:image2}
    \end{subfigure}
    \hfill
    \begin{subfigure}[b]{0.22\textwidth}
        \centering
        \includegraphics[width=\textwidth]{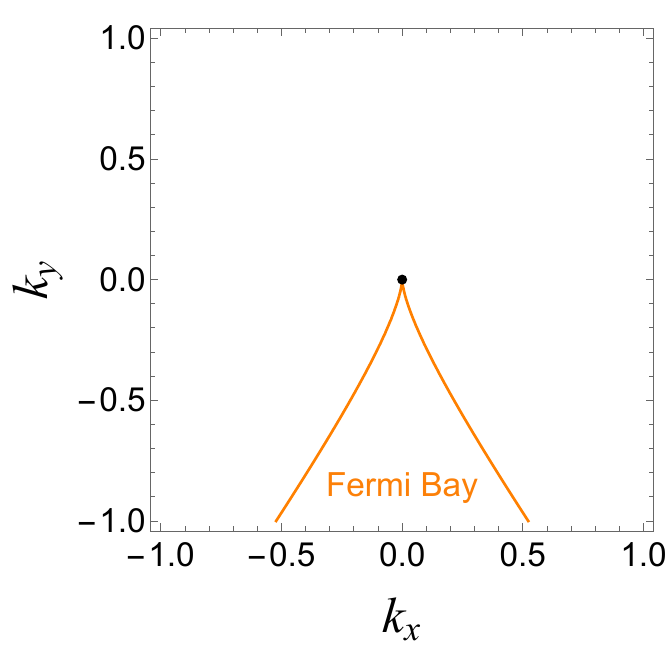}
        \caption{Tilt  $\zeta=1$}
        \label{fig:image3}
    \end{subfigure}
      \hfill
    \begin{subfigure}[b]{0.22\textwidth}
        \centering
        \includegraphics[width=\textwidth]{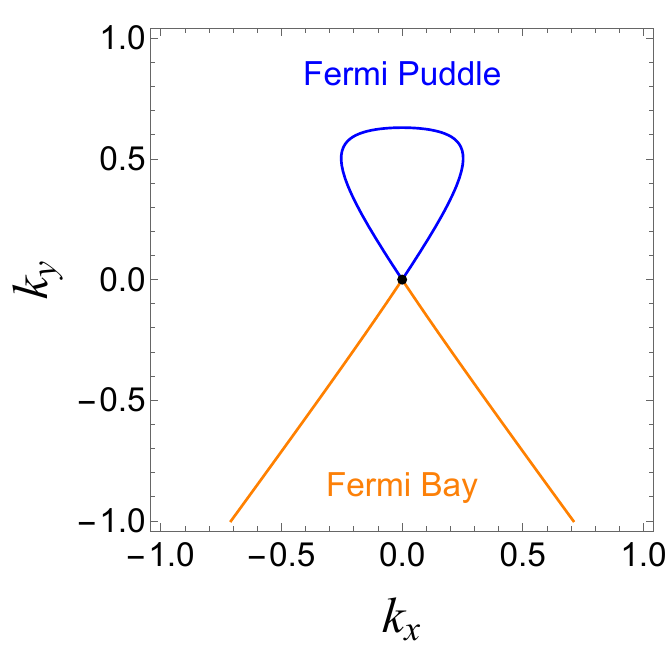}
        \caption{Tilt  $\zeta=6/5$}
        \label{fig:image3}
    \end{subfigure}
    \caption{ The quadratic dispersion relation of the tilted Dirac cones. Top panel: dispersion relations with the orange upper branch, blue lower branch and the green zero-energy plane. Bottom panel: zero-energy Fermi surface for different tilt parameter. (a) Tilt $ \zeta=0$, Dirac cone with a dot-like zero-energy Fermi surface. (b) Tilt $ \zeta=6/7$, Dirac cone whose zero-energy Fermi surface is a dot and Fermi bay.  (c) Tilt $ \zeta=1$, Dirac cone with the zero-energy Fermi bay just touches the Dirac point. (d)  Tilt $ \zeta=6/5$, the Fermi Bay (orange curve) touches the Dirac point, and a  Fermi puddle (blue closed curve) forms.}  
    \label{quadratic dispersion relation}
\end{figure}

We show the quadratic dispersion of the tilted Dirac cones in Figure \ref{quadratic dispersion relation}. When $\zeta=0$, the zero-energy Fermi surface is a single point (the Dirac point). When  $\zeta>0$,  the orange curve in the lower panel of Figure \ref{quadratic dispersion relation}, which we refer to as the {\it Fermi bay}, emerges. The Fermi Bay appears far from the Dirac point and gradually moves closer to the Dirac point as the tilt parameter $\zeta$ increases. When $\zeta=1$, the Fermi bay touches the Dirac point. Beyond this critical point, as the tilt parameter $\zeta$ increases, i.e., $\zeta>1$, a blue closed curve in the lower panel of Figure \ref{quadratic dispersion relation}, which we refer to as the {\it Fermi puddle},  begins to appear. As $\zeta$ increases further, the size of the Fermi puddle gradually grows larger.

\section{Entropy of tilted Dirac cone material}\label{sec linear}
Since the emergent spacetime metric associated with the tilted dispersion hosts a horizon at $\zeta=1$, it is natural to ask whether this analogue geometry also reproduces the thermodynamic properties of black holes—most notably, the Bekenstein–Hawking entropy \cite{Bekenstein:1972tm,Bekenstein,Hawking:1975vcx}. The key idea is that the tilt-induced modification of the energy spectrum alters the density of states, which directly influences the entropy. To address this, we examine how the entropy of Dirac cone material compares with the Bekenstein-Hawking entropy of a $2+1$-dimensional black hole. As we proposed in the Introduction, we consider the analogue black holes that match this entropy, ``smart holes''. In this section, we start by studying the linear dispersion relation with a space-dependent  tilt parameter, and compute the associated entropy density of the tilted Dirac cone in the condensed matter system.

For the tilted Dirac cone materials, with continuous energy spectrum, the entropy density is given by

\begin{align}
    s_1&=\frac{2}{(2\pi)^2} \int d^2\boldsymbol{p}\left(\frac{\gamma\beta_{\rm prop} E}{e^{\gamma\beta_{\rm prop} E}+1}+\ln{(1+e^{-\gamma \beta_{\rm prop} E})}\right).
\end{align}
This entropy density formula is derived from the standard statistical mechanics approach, where the factor of 2 accounts for the spin degeneracy of the fermions. The dispersion relation is represented by \( E \), while the Fermi-Dirac distribution is modified by a Lorentz-like factor \( \gamma = (1 - \zeta^2)^{-1 / 2} \) \cite{Moradpouri:2022mhi}, which depends on the tilt parameter \( \zeta \) in the emergent spacetime metric and  functions analogously to a redshift factor  in curved spacetime, and $\beta_{\rm prop}=1/T_{\rm prop}$ is inverse  temperature, according to an analogue static observer. In principle, a nonzero chemical potential term may be added to this formula. However, particle-hole symmetry, reflected in the identical area or shape of the particle and hole Fermi puddles, implies the conservation of the total particle number. When this symmetry is present, the chemical potential can be set to zero, corresponding to the energy at the Dirac points. We introduce a new parameter $\theta'=\theta-\frac{\pi}{2}$, with which the linear energy spectrum for tilted  Dirac cone materials given by equation (\ref{linearE}) is read as
\begin{align}
    E=(s+\zeta\cos{\theta'})v_F p.
\end{align}

If we note $\beta$ as the inverse laboratory temperature, the relationship between the proper temperature and the laboratory temperature is given by $  \beta=\gamma\beta_{\rm prop}$ \cite{Tolman:1930zza,Tolman:1930ona}. Then, the entropy density is given by
\begin{align}\label{ss1}
    s_1&=\frac{2}{(2\pi)^2}\int_0^{2\pi}d\theta' \int_{p_{min}}^{p_{max}}pdp \left(\frac{\beta E}{e^{\beta E}+1}+\ln{(1+e^{- \beta E})}\right).
\end{align}
Let us change the integration variable from momentum $p$ to energy $E$ and apply a new variable $E'=\beta E$. Then, the entropy density is partitioned into two parts; the angle integration over $\theta'$ and the integral over energy $E'$
\begin{align}
    s_1&=\frac{2}{(2\pi)^2}\int_0^{2\pi} \frac{1}{v_F^2(s+\zeta\cos{\theta'})^2}d\theta' \int_{E_{min}}^{E_{max}}EdE \left(\frac{\beta E}{e^{\beta E}+1}+\ln{(1+e^{- \beta E})}\right)\nn\\
     &= \frac{2}{(2\pi)^2v_F^2\beta^2}\int_0^{2\pi} \frac{1}{(s+\zeta\cos{\theta'})^2 }d\theta' \int_{E'_{min}}^{E'_{max}}E'dE'  \left(\frac{E'}{e^{E'}+1}+\ln{(1+e^{-E'})}\right).
\end{align}
Taking into account the sum of two Energy branches $s=\pm 1$, and considering both spin-up and spin-down fermions, we find
\begin{align}\label{enden}
      s_2&=4s_1= \frac{16}{(2\pi)^2v_F^2\beta^2}\int_{0}^{\pi} \frac{1}{(1+\zeta\cos{\theta'})^2 }d\theta' \int_{0}^{E'_{max}}E'dE'  \left(\frac{E'}{e^{E'}+1}+\ln{(1+e^{-E'})}\right).
\end{align}
In tilted Dirac cone materials, the tilt parameter \( \zeta \) modifies the energy spectrum and alters the density of states. Consequently, the entropy density becomes a function of \( \zeta \), reflecting the impact of tilt on the thermodynamic behavior of the system. Moreover,  the entropy density can acquire a spatial dependence if the tilt parameter is not fixed but varies spatially.   This assumption arises naturally in the construction of a black hole analogue, where the light cone gradually tilts as one approaches the event horizon. To capture this behavior in the tilted Dirac cone material, we model the tilt parameter as a spatially varying function, rather than a constant. To simplify this construction, we consider the case where $\boldsymbol{\zeta}$  has no  $x$-component, and $\zeta_y$ depends solely on the coordinate  $y$.  In this setup, based on the emergent spacetime metric in equation~(\ref{metric}), the event horizon is located at $\zeta_y(y) = 1$ when considering a hypersurface of constant $y$. The associated surface gravity is then explicitly given by
\begin{equation}
\kappa = v_F^2|   \zeta_y(y) \zeta_y'(y)|\big|_{y=y_{\text{horizon}}}= v_F^2|  \zeta_y'(y)|\big|_{y=y_{\text{horizon}}}.
\end{equation}
This expression reveals that the surface gravity arising from the emergent spacetime metric~(\ref{metric}) is fundamentally governed by the spatial gradient of the tilt parameter. Consequently, in order to introduce the notion of surface gravity—which manifests as the temperature in black hole spacetimes—it is essential to allow the tilt parameter $\boldsymbol{\zeta}$ to vary spatially in the tilted Dirac cone system. 
Hence, a natural and physically intuitive scenario in materials exhibiting tilted Dirac cones is to assume a linear spatial dependence for the tilt parameter. Specifically, the tilt parameter $\boldsymbol{\zeta}$ is given by
\begin{align}
    \zeta_x=0, \qquad \zeta_y(y)= -\frac{\kappa}{v_F^2}y + 1,
\end{align} where the spatial gradient of the tilt parameter plays the role of  surface gravity of a black hole, which is consistent with the one obtained from the emergent spacetime metric (\ref{metric}). Within our tilted Dirac cone system, $\zeta=1$, where $y=0$, precisely corresponds to the analogue event horizon. Moreover, at $y=\frac{v_F^2}{\kappa}$, the tilt parameter $\zeta$ vanishes, clearly marking the boundary of the analogue ``outside horizon" region,  characterized by $0<\zeta<1$. This effectively models the black hole exterior region, offering an experimentally accessible platform for investigating black hole-like phenomena in condensed matter systems.  In contrast, the region where $\zeta>1$ corresponds to the analogue black hole interior, enabling further exploration of black hole interior physics in a controlled laboratory setting.

Finally, to compute the total entropy, we integrate the entropy density over the spatial coordinates $x$ and $y$. We consider a two-dimensional sample with tilted Dirac cone with length $L_x$ and width $L_y=\frac{v_F^2}{\kappa}\zeta_{max}$. Then, the total entropy is given by
\begin{align}
    S=\int s_2 dxdy=\frac{v_F^2 L_x}{\kappa}\int d\zeta s_2,
\end{align}
which, applying \eqref{enden}, is computed as
\begin{align}\label{total}
    S 
    &=\frac{16 L_x}{(2\pi)^2\kappa\beta^2}\int_{{\zeta_{min}}}^{\zeta_{max}} d\zeta\int_0^{\pi}d\theta'  \frac{1}{(1+\zeta\cos{\theta'})^2 }\int_{E'_{min}}^{E'_{max}}E'dE'  \left(\frac{E'}{e^{E'}+1}+\ln{(1+e^{-E'})}\right).
\end{align}
We first compute the energy integral to find
\begin{align}
     F[E'] &\equiv \int^{E'} E''dE''  \left(\frac{E''}{e^{E''}+1}+\ln{(1+e^{-E''})}\right)\nn\\
    &=    -E'^2 \log[1 + e^{-E'}] + 3 E'\rm{Li}_2(-e^{-E'}) + 3 \rm{Li}_3 (-e^{-E'}),
\end{align}
where $F[0]=-\frac{9\zeta_{\rm R}(3)}{4}$. We note that $\zeta_{\rm R}(z)$ here is the Riemann zeta function, and $\rm{Li}_s(z)$ is the polylogarithm function of order $\rm{s}$ and argument $\rm{z}$, which is defined as the series
\begin{align}
    {\displaystyle \rm{Li}_s(z)=\sum _{k=1}^{\infty }{z^{k} \over k^{s}}=z+{z^{2 } \over 2^{s}}+{z^{3} \over 3^{s}}+\cdots }.
\end{align}

If we assume that $E'_{max}$ is infinitely large and $ E'_{min}$ is zero, then the entropy density  equation (\ref{enden})  becomes
\begin{align}\label{entropy density}
    s_2=\frac{16}{(2\pi)^2v_F^2\beta^2} \frac{\pi}{(1-\zeta^2)^{3/2}}\times \frac{9\zeta_{\rm R}(3)}{4}=\frac{9\zeta_{\rm R}(3)}{\pi}\frac{T^2}{v_F^2 (1-\zeta^2)^{3/2}}.
\end{align}
This result is in agreement with \cite{Moradpouri:2022zwa} from the holographic approach. In the upright Dirac cone limit, where the tilt parameter vanishes ($\zeta = 0$) in equation (\ref{entropy density}), we also recover the result presented in \cite{PhysRevLett.103.025301}. In the analogue black hole horizon limit ($\zeta \to 1$), the entropy density diverges. Consequently, for linear dispersion, the total entropy integral as given in equation (\ref{total}), becomes infinite. To regularize this integral, one must either introduce a cutoff on $E'_{max}$ or incorporate non-linear corrections.

\section{Analogue black holes with the right
temperature and entropy}\label{sec num}
\begin{figure}[!h]
    \centering
    % 子图1
    \begin{subfigure}[b]{0.45\textwidth}
        \includegraphics[width=\textwidth]{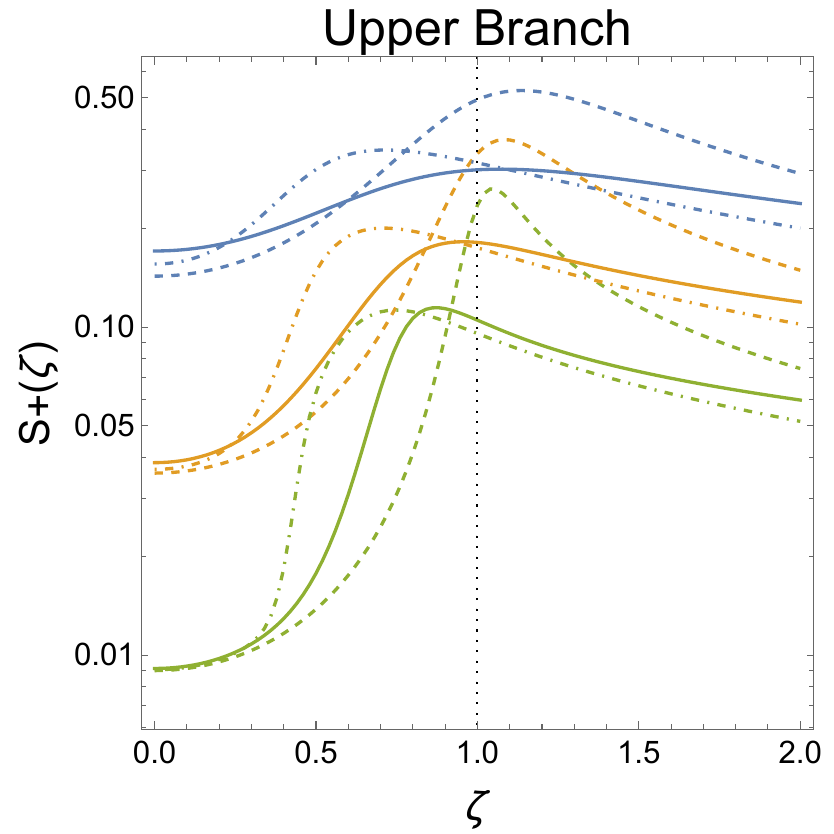}
        \caption{Entropy for the upper branch.}
        \label{upper}
    \end{subfigure}
    % 子图2
    \begin{subfigure}[b]{0.45\textwidth}
        \includegraphics[width=\textwidth]{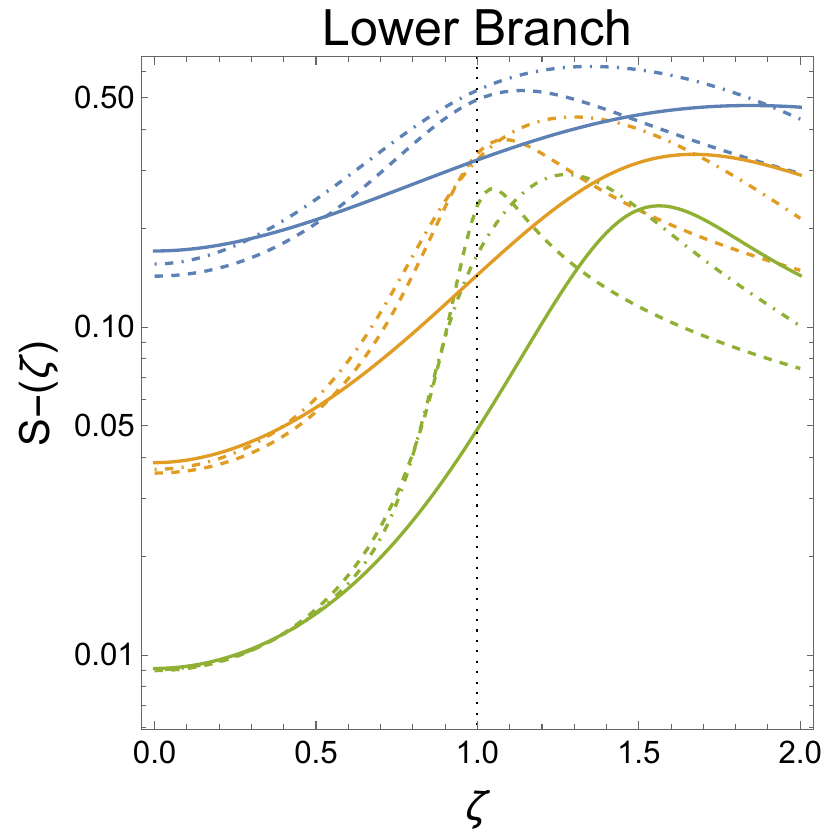}
        \caption{Entropy for the lower branch.}
        \label{lower}
    \end{subfigure}
    \caption{Entropy density as a function of the tilt parameter \( \zeta \). The blue, orange, and green curves correspond to inverse temperatures \( \beta = 4 \), \( \beta = 8 \), and \( \beta = 16 \), respectively. The dashed lines represent results using the linear dispersion relation, the dot-dashed lines use the quadratic dispersion relation, and the solid lines are based on the full band structure.}
\end{figure}

In this section, we numerically compute the entropy density for the upper and lower branches as a function of the tilt parameter $\zeta$, and show them in Figures \ref{upper} and \ref{lower}. The numerical entropy density we compute is given by twice that in equation (\ref{ss1}), since we are considering both spin-up and spin-down fermions. Note that the integration region we choose in the reciprocal space is half of the Brillouin zone because the (first) Brillouin zone contains two Dirac cones.  To be specific, in the linear and quadratic cases, we take a circular integration region with radius $\Delta |\boldsymbol{k}|=\sqrt{\frac{4 \pi }{\sqrt{3}}}\frac{1}{a}$. This can be regarded as 
 a cutoff on momentum or energy.  
 
 In Figures \ref{upper} and \ref{lower}, the dashed lines show results for the linear dispersion relation, while the dot-dashed lines include quadratic corrections. The solid lines provide a more accurate depiction of the system by accounting for the full band structure, which includes higher-order terms that go beyond both the linear and quadratic approximations. Each color represents an inverse temperature value, the blue, orange, and green curves represent inverse temperatures \( \beta = 4 \), \( \beta = 8 \), and \( \beta = 16 \), respectively.
Across all temperature ranges, the entropy initially increases with $\zeta$, peaks around $\zeta \approx 1$, and then gradually decreases as $\zeta$ increases further. As the tilt parameter changes, the system's entropy demonstrates a significant dependence on this parameter, especially when considering more complex dispersion relations, which affect both the peak values and their positions. The exact height and shape of these peaks depend on the inverse temperature and the choice of the dispersion relation.  At lower temperatures (i.e., higher $\beta $), the entropy decreases and the curves show sharper peaks. In contrast, at higher temperatures (i.e., lower $\beta$), the entropy increases and the peaks become broader.

As shown in Figure \ref{upper}, for a fixed temperature (that is, the same color of the curve), the peak of quadratic dispersions (dot-dashed) shifts to a slightly lower value of
$\zeta$ compared to the linear case (dashed). This shift can be attributed to the contribution of the orange Fermi curve (Fermi Bay) shown in Figure \ref{quadratic dispersion relation}, which appears before $\zeta=1$. The peak location is pushed to the right when we include higher-order terms in the quadratic dispersion relation. In contrast, as shown in Figure \ref{lower}, the peak for  quadratic dispersion relation shifts to a slightly larger value of $\zeta$ compared to the linear case, for the lower branch. This indicates that incorporating quadratic corrections delays the point at which the system reaches its maximum entropy. This comes from the contribution of Fermi puddle (blue curve in Figure \ref{quadratic dispersion relation}), which appears after $\zeta=1$. The Fermi puddle emerges as a distinct pocket structure in the zero-energy Fermi surface beyond the critical tilt, fundamentally reshaping the entropy density distribution. This structure modifies the density of states, effectively creating a reservoir of low-energy excitations. These localized states play a crucial role in enhancing entropy, particularly in the lower energy branch, where the entropy density peaks in the vicinity of the Fermi puddle. The peak location is pushed further to the right when including higher-order terms in the dispersion relation. For both branches, incorporating quadratic dispersion corrections affects the tilt parameter $\zeta$ at which entropy density peaks. In the upper (lower) branch, this causes the peak to shift to smaller (larger) $\zeta$.   When the full-band structure effects are included, the peaks are always pushed to the right (i.e., higher $\zeta$) compared to the quadratic case, reflecting the more complex behavior of the system's entropy as a function of $\zeta$. One may be tempted to interpret these trends in terms of the finite-temperature renormalized theory, but we shall defer such interpretation to future work.

\begin{figure}[!h]
    \centering
\includegraphics[width=0.7\textwidth]{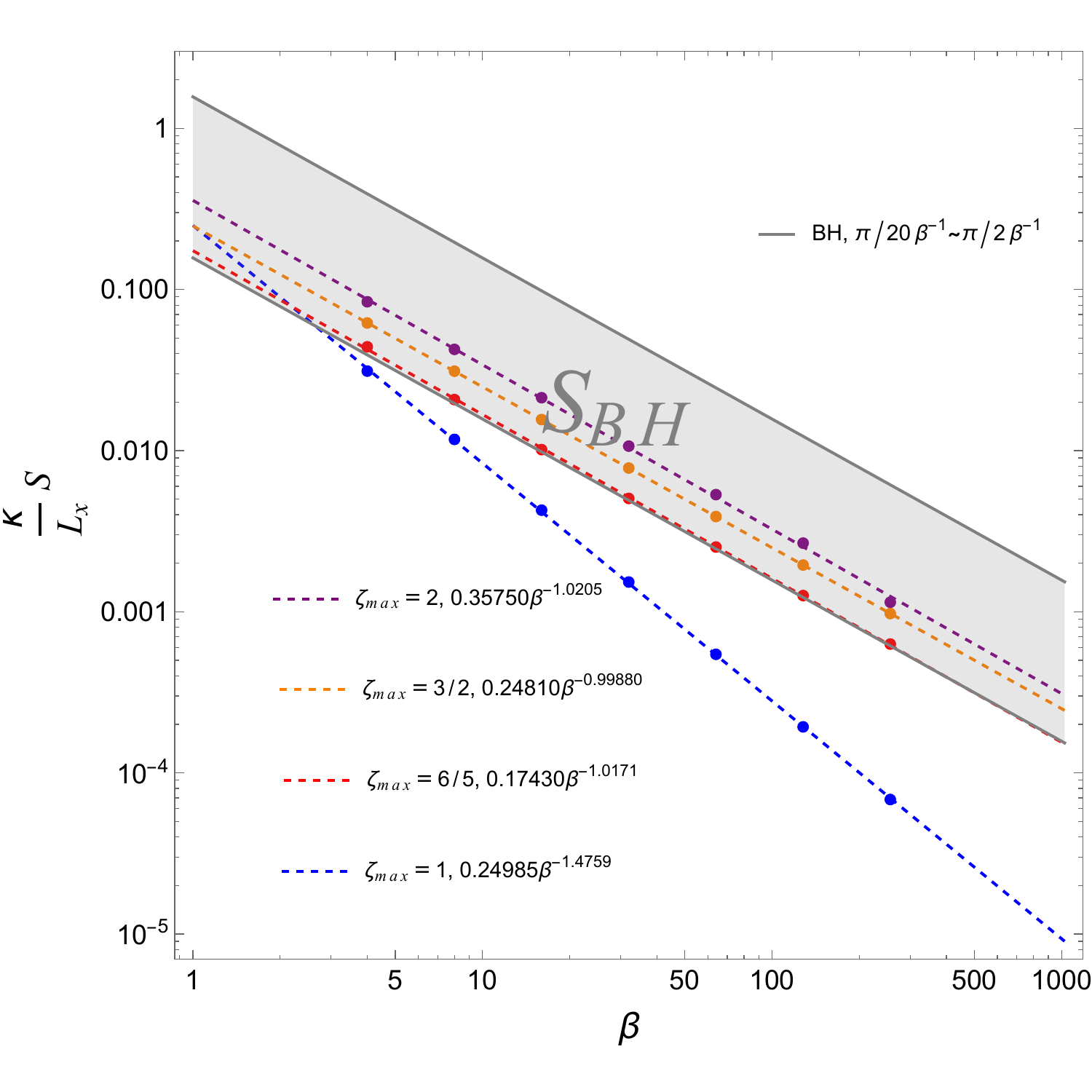}
    \caption{ The relationship between the total entropy (times surface gravity) of the lower branch for the linear dispersion  (\ref{linearE}) and the inverse temperature, $\beta$, where the integral over the tilt parameter ranges from 0 to $\zeta_{\text{max}}$. The blue, red, orange, and purple points represent $\zeta_{\text{max}} = 1, \frac{6}{5}, \frac{3}{2}$, and $2$, respectively. The lines represent the best-fit power laws. The shaded region shows the expected range of entropy times surface gravity of a black hole, which is proportional to the temperature (see the text).}
    \label{s vs beta linear}
\end{figure}

\begin{figure}[!h]
    \centering
\includegraphics[width=0.7\textwidth]{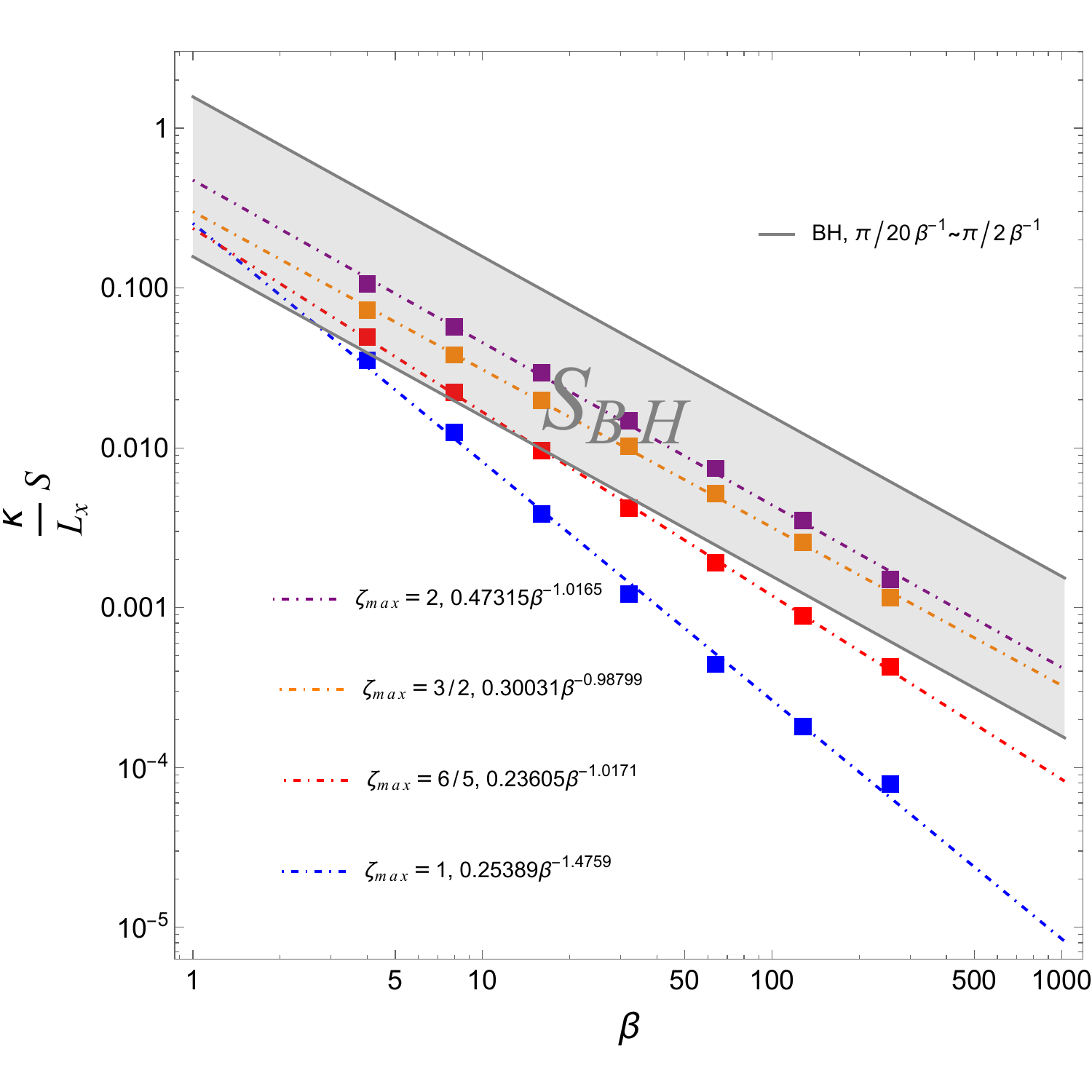}
    \caption{Same as Figure \ref{s vs beta linear}, but with the quadratic dispersion relation (\ref{quadratic}).}
    \label{s vs beta quadratic}
\end{figure}

\begin{figure}[!h]
    \centering
\includegraphics[width=0.7\textwidth]{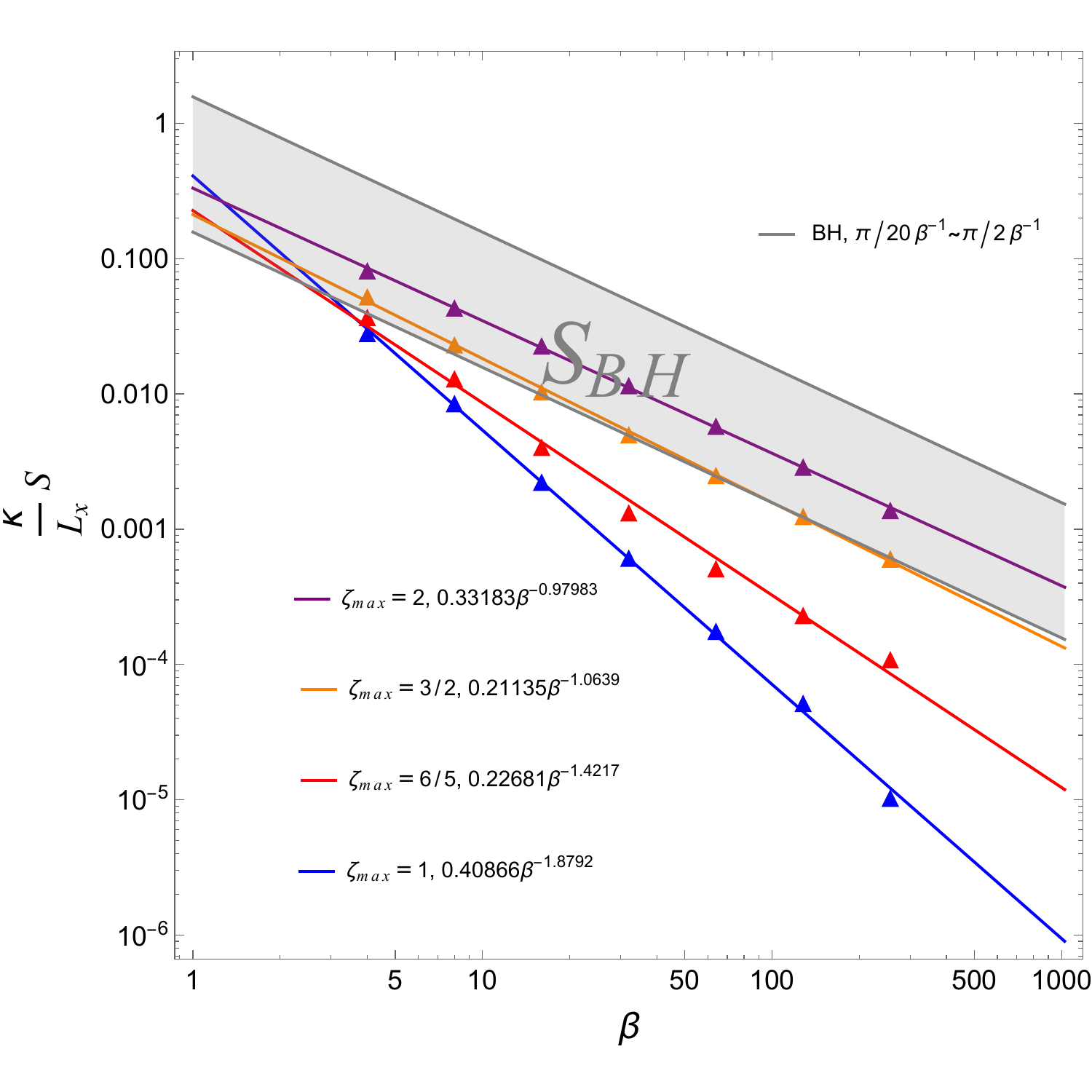}
    \caption{Same as Figure \ref{s vs beta linear}, but with the full nonlinear dispersion relation (\ref{full}).}
    \label{s vs beta full}
\end{figure}

Figures \ref{s vs beta linear}, \ref{s vs beta quadratic},  and \ref{s vs beta full} show the total entropy (after integrating over the $\zeta$ dependence) of the lower (dominant) branch, as a function of the inverse temperature $\beta$. The best-fit power-law dependencies of entropy vs. inverse temperature are also shown, which appear as lines in the log-log plot.  As $\beta$ becomes too large, the integrand becomes very sharply peaked (and vanishingly small), thus we should not trust the numerics in this region. When $\beta$ is small, i.e., temperature is high, excitations go far from the Fermi surface, and thus a power-law temperature dependence is not expected.

Let us first focus on Figure \ref{s vs beta linear}, i.e., the case of linear dispersion. When $\zeta_{\text{max}} = 1$, the best-fit power-law for total entropy is approximately 
\begin{align}
    S\propto \frac{L_x}{\kappa} \frac{p_{max}^{1/2}}{
\beta^{3/2}},
\end{align}
However, most interesting is the fact that when $\zeta_{\rm max}$ exceeds 1, i.e., we include entropy inside the horizon, the best-fit total entropy is approximately linearly dependent on  $p_{max} T$: 
\begin{align}
    S\propto\frac{L_x}{\kappa} \frac{p_{max}}{
\beta},
\end{align}

This is exactly what we expect from Bekenstein-Hawking entropy for a black hole horizon, if temperature is related to surface gravity, i.e., $2\pi \kappa \beta =1$, {\it à la} Hawking \cite{Hawking:1974rv,Hawking:1975vcx}.

In Figures \ref{s vs beta quadratic} and \ref{s vs beta full}, the total entropy times surface gravity for both the quadratic and full dispersion relations also exhibit a linear dependence on $p_{\text{max}} T$ at a critical value of $\zeta_{\rm max}$ greater than 1. This is consistent with the observed phenomenon where the entropy density peak shifts to the right in Figure \ref{lower}, suggesting that the effective horizon forms after $\zeta = 1$.

From the above numerical computations, we conclude that the total entropy for 2d tilted Dirac cone material  with length $L_x$ and width $L_y=\frac{v_F^2}{\kappa}\zeta_{max}$ is 
\begin{align}
    S_{DC}
    &\approx \eta\frac{L_x p_{max}T}{ \kappa},
\end{align}
where $\eta$ is a dimensionless factor, depending on the $\zeta_{max}$ value we choose. The larger $\zeta_{max}$ we consider, the larger the factor $\eta$. Let us compare the above result with the Bekenstein-Hawking entropy of $2+1$ dimensional Ba{ñ}ados, Teitelboim, and Zanelli (BTZ) black hole \cite{Banados:1992wn} with horizon radius $r_+$, which is given by one fourth of the horizon size \cite{Carlip:1995qv} 
\begin{align}
    S_{BH}=\frac{2\pi r_+}{4G},
\end{align}

In Figures \ref{s vs beta linear}, \ref{s vs beta quadratic} and \ref{s vs beta full}, we choose Newton's constant (= 2+1d Planck length) to be in the range $a \leq G\leq 10 a$, in terms of lattice spacing $a$, to plot the shaded region, representing the black hole region. 
If we make the assumption that the temperature $T$ is interpreted as the analogue black hole surface gravity $\kappa$, the momentum cutoff $p_{max}$  is the inverse of the Newton constant $G$, and the length scale $L_x$ is interpreted as the analogue black hole horizon size $2\pi r_+$,
\begin{align}
\begin{cases}
T=\frac{\kappa}{2\pi}\\
p_{max}=\frac{\pi}{2\eta}\frac{1}{G}\\
L_x=2\pi r_+
\end{cases},
\end{align}
then an intriguing parallel can be established.
{\it The thermal entropy of the analogue black hole constructed by tilted Dirac cone materials matches the Bekenstein-Hawking entropy of a real BTZ black hole system, as we had desired for our ``smart hole''}. This correspondence not only sheds light on the behaviors of both condensed matter and gravitational systems, but also paves the way for novel explorations into the intricate interplay between quantum theory, gravity, and thermodynamics.

At last, let us provide a justification for the assumed relationship between the temperature and the surface gravity of analogue black holes in condensed matter systems, matching that of Hawking temperature \cite{Hawking:1974rv,Hawking:1975vcx}. If we introduce a symmetric (Hermitian) Hamiltonian $H_{sym}$,
\begin{align}
    H_{\rm sym}=v_F
    \begin{bmatrix}
   ( \boldsymbol{p}\cdot \boldsymbol{\zeta}+ \boldsymbol{\zeta}\cdot \boldsymbol{p})/2&p_x\mp ip_y\\
    p_x\pm ip_y&   ( \boldsymbol{p}\cdot \boldsymbol{\zeta}+ \boldsymbol{\zeta}\cdot \boldsymbol{p})/2
    \end{bmatrix},
\end{align}
then the original free Hamiltonian $H_0$ given in equation (\ref{ham1}) and the symmetric one $H_{\rm sym}$ is related by
\begin{align}
    H_0=H_{\rm sym}+\frac{\kappa}{2}i.
\end{align}
We now examine how the relation between surface gravity and temperature in the analogue black hole can arise from the Kubo–Martin–Schwinger (KMS) condition \cite{Kubo:1957mj,Martin:1959jp,Haag:1967sg}, a fundamental criterion for thermal equilibrium in quantum field theory \cite{le2000thermal,laine2016basics,Mustafa:2022got,yang2011introduction}. To this end, we consider the thermal correlation function of two quantum field operators \begin{align}
\langle \psi(\mathbf{x}_1, t) \psi(\mathbf{x}_2, 0) \rangle_\beta 
&= \frac{1}{Z} \mathrm{Tr} \left[ e^{-\beta H_0} \psi(\mathbf{x}_1, t) \psi(\mathbf{x}_2, 0) \right] \nonumber\\
&= \frac{1}{Z} \mathrm{Tr} \left[ \psi(\mathbf{x}_1, t) e^{-\beta H_{\text{sym}}} e^{\beta H_{\text{sym}}^\dagger} \psi(\mathbf{x}_2, 0) e^{-\beta H_0} \right] \nonumber\\
&= \frac{1}{Z} \mathrm{Tr} \left[ \psi(\mathbf{x}_1, t) e^{-\beta H_{0}} e^{\beta\kappa i/2} e^{\beta\kappa i/2}e^{i(-i\beta H_{0}^\dagger)} \psi(\mathbf{x}_2, 0) e^{-i(-i\beta H_{0})} \right] \nonumber\\
&= \frac{1}{Z} \mathrm{Tr} \left[ \psi(\mathbf{x}_1, t) e^{-\beta H_0} \psi(\mathbf{x}_2, -i\beta) e^{\beta\kappa i} \right] \nonumber\\
&= \frac{1}{Z} \mathrm{Tr} \left[ e^{-\beta H_0} \psi(\mathbf{x}_2, -i\beta) \psi(\mathbf{x}_1, t) e^{\beta\kappa i} \right]\nonumber \\
&= \langle \psi(\mathbf{x}_2, -i\beta) \psi(\mathbf{x}_1, t)e^{\beta\kappa i}  \rangle_\beta, \label{periodic}
\end{align}
 where we used the cyclic property of the trace,  $Z$ is the partition function, and $\beta$ is the inverse temperature.  
The KMS condition $\langle \psi(\mathbf{x}_1, t) \psi(\mathbf{x}_2, 0) \rangle_\beta = \langle \psi(\mathbf{x}_2, -i\beta) \psi(\mathbf{x}_1, t)  \rangle_\beta\nonumber$ immediately yields the identification of the temperature with the surface gravity
\begin{align}
    \beta = \frac{2\pi}{\kappa} \quad \Rightarrow \quad T = \frac{\kappa}{2\pi}. \label{Hawking_temp}
\end{align}
This establishes a direct connection between the thermal properties of the quantum field and the analogue gravitational background. Moreover, the KMS condition will lead to periodic boundary conditions for bosonic fields and anti-periodic boundary conditions for fermionic fields in imaginary time
due to the commutation and anti-commutation relations, respectively. Importantly, the temperature obtained from the KMS condition precisely matches the characteristic Hawking temperature for a real black hole \cite{Hawking:1974rv,Hawking:1975vcx}, providing strong evidence that the quantum fields in our analogue black hole setup behave as if it were in an Unruh-like thermal state \cite{Unruh76,Dappiaggi:2009fx}. This, in turn, supports the robustness of Hawking-like radiation in our condensed matter realization. Further insight into the origin of this Hawking-like radiation is provided by a proposed microscopic mechanism \cite{volovik2016black}. This mechanism suggests that the critically tilted Dirac cone, marking the formation of an analogue horizon, triggers transitions between energy branches and enables particle–hole pair creation. Hawking-like radiation is then simulated via such particle–hole pair creation across the horizon, driven by an initial imbalance between occupied and unoccupied energy states. The corresponding Hawking temperature of this analogue system is governed by the spatial gradient of the tilt parameter at the horizon, reflecting the role of surface gravity in curved spacetime.
As further confirmation, one could analyze the particle number spectrum for evidence of a thermal distribution \cite{Hawking:1974rv,Hawking:1975vcx}, or couple an Unruh–DeWitt detector~\cite{Unruh76, DeWitt80} to probe the thermal behavior of its response. We leave these investigations for future work.

At this stage, a natural question arises: what justifies setting equation (\ref{Hawking_temp}) as the temperature of the tilted Dirac cone material in the laboratory, as opposed to say room temperature? At face value, the reasoning could be that equation (\ref{periodic}) dictates that (\ref{Hawking_temp}) is the only temperature ensuring a well-defined, Hermitian thermal continuum limit for the tilted Dirac cone material. If this interpretation holds, it would parallel the requirement for quantum fields near a black hole horizon to adopt the Hawking temperature in order to maintain a Hadamard state (to ensure a finite renormalized stress-energy tensor \cite{PhysRevD.21.2185}, or a  finite 1-loop backreaction \cite{Mathur:2024mvo}). In the following, we present a more detailed discussion of the validity of the temperature–surface gravity relation from multiple perspectives.

Indeed, as one might point out, there is no guarantee that the temperature set by the cryogenic refrigerator in the laboratory satisfies  \( T = \kappa/2\pi \),   which is determined by the varying tilt parameter in our Dirac cone material. However, as we showed in \eqref{periodic}, if this condition is not satisfied, the continuum theory does not obey the thermal statistics. In other words, the low energy modes, far from the horizon, become blueshifted as they travel towards the horizon, and leave the regime of validity of the continuum theory. At an intuitive level, the condition $T=\kappa/ 2\pi$ ensures that this blueshift is precisely compensated by the dissipation caused by the non-unitarity of the Hamiltonian $H_0$. Otherwise, the system will not have a well-defined continuum (or infrared) description, as most of the degrees of freedom live near the lattice scale, in the vicinity of the horizon.

The temperature in the entropy density expressions indeed corresponds to the laboratory temperature that governs the thermal occupation of energy states in the condensed matter system. At first sight, this may appear unrelated to the Hawking temperature, which arises from the surface gravity of a gravitational horizon. While this distinction may seem to suggest a discrepancy, it in fact highlights a key feature of our analogue black hole paradigm. In gravitational systems, Hawking temperature dictates the thermal behavior of quantum fields in curved spacetime. In condensed matter systems, the laboratory temperature plays an analogous role, governing the thermal occupation of modes in the emergent geometry. The bridge between the two is established by the emergence of an analogue horizon as the Dirac cone undergoes critical tilting. One can thus tune the laboratory temperature to match the emergent Hawking temperature determined by the spatial gradient of the tilt parameter 
\begin{equation}
T = \frac{\hbar v_F}{2\pi k_B} \left| \, \frac{d\zeta}{dy} \right|.
\end{equation}
This matching is not trivial---it defines the regime in which the analogue system not only mimics the geometric structure of a black hole but also reproduces its thermodynamic behavior. In this regime, the statistical entropy from the Fermi--Dirac distribution becomes directly comparable to the Bekenstein--Hawking entropy of the analogue black hole. This thermodynamic consistency between the laboratory temperature and the emergent geometry offers a physically motivated route toward realizing black hole thermodynamics \cite{Bardeen:1973gs,Bekenstein:1972tm,Bekenstein} in condensed matter systems.  

 While black hole thermodynamics provides deep theoretical insights, its experimental realization in astrophysical contexts remains out of reach. For instance, the Hawking temperature of a solar-mass black hole is below \(10^{-6}\,\text{K}\) \cite{Hawking:1974rv}, a value far beneath the cosmic microwave background and inaccessible to any current observational technique. By contrast, the analogue Hawking temperature in our model depends on controllable material parameters. For example, with a typical Fermi velocity \(v_F \sim 10^6\,\text{m/s}\) \cite{hwang2012fermi,lima2015controlling} and a moderate tilt gradient \(\frac{d\zeta}{dy} \sim 10^6\,\text{m}^{-1}\)— which corresponds to a unit change in the tilt parameter over \(  10000\,\text{\AA} \)— the emergent temperature can reach a few kelvin. For steeper gradients, such as \(\frac{d\zeta}{dy} \sim 10^8\,\text{m}^{-1}\), where the tilt changes by order one over just \(100\,\text{\AA}\), the temperature can be engineered to exceed room temperature. This highlights the experimental accessibility of our model and suggests that black hole thermodynamics, traditionally confined to inaccessible regimes, can now be probed in a table-top experiment.

Finally, let us briefly discuss the scale separation in our model. The modified dispersion relations (MDR) corrections originate from the underlying lattice structure and become significant only at short length scales of order \( L_{\text{MDR}} \sim a \) (where \( a \) is the lattice spacing), corresponding to an energy scale \( E_{\text{MDR}} \sim 1 / a \). Meanwhile, the variation of the emergent spacetime background—characterized by the surface gravity \( \kappa \) of the analogue black hole—occurs over much larger length scales. In our setup, we explicitly engineer the background to vary over \( L_\kappa \sim 10^2a \) (corresponding to a Hawking temperature on the order of room temperature), and in some cases over \( 10^4a \) (yielding temperatures of a few kelvin), which corresponds to an energy scale \( E_\kappa \sim 10^{-2}a^{-1} \) or smaller. This immediately yields a clear hierarchy
\[
E_\kappa \ll E_{\text{MDR}}.
\]
This separation of scales is essential for the robustness of Hawking-like radiation, as emphasized in Refs.~\cite{barcelo2011analogue,Unruh:2004zk}. By ensuring that \( E_\kappa \ll E_{\text{MDR}} \), our system guarantees that the relevant modes involved in the Hawking-like process remain well in the relativistic (linear) regime of the dispersion relation. Therefore, we believe that the Hawking-like effect in our system remains robust for modified dispersion relations.
\section{Closing remarks}\label{sec con}

From the condensed matter perspective, the nearest-neighbor hopping is responsible for the formation of upright Dirac cones, and the anisotropic next-nearest-neighbor hopping tilts the Dirac cones. Different materials differ in details of additional couplings at the lattice scale $a$. Such differences only modify the higher energy parts of the bands, but does not alter the Dirac part. Therefore the Dirac theory is the  universal theory of this class of materials. For the tilted Dirac cone band structure, including the non-linear corrections to the dispersion relation changes the geometry of the zero-energy Fermi surface. When $0<\zeta< 1$, we observe the appearance of a {\it Fermi bay} disconnected from Dirac points. When $\zeta=1$, the Fermi bay touches the Dirac points, while for $\zeta\gtrsim 1$, we observe the appearance of a {\it Fermi puddle} on the opposite side. Different tilted Dirac materials can differ in the size and shape of their Fermi puddles.

We find that the entropy density rises with the tilt $\zeta$ initially, peaks around $\zeta= 1$, and then gradually declines as $\zeta$ increases further. When the second or higher order dispersions are considered, the peak moves to larger values of $\zeta$.
We also compute the total entropy-temperature relation, up to different maximum tilts, for the three different dispersions. In conclusion, our study demonstrates that tilted Dirac cones in 2D Dirac materials, particularly borophene, can effectively serve as analogues for black hole horizons, capturing both the temperature and the entropy characteristics of black holes. The entropy of this condensed matter system mirrors the Bekenstein-Hawking entropy, and the peak is situated behind the event horizon of the analogue black hole. This novel analogue, which we call the smart hole, highlights the potential for condensed matter systems to model black hole thermodynamics (and dynamics) in laboratory environments, offering valuable insights into the quantum aspects of black hole physics.

\acknowledgments

J.Y. is grateful for support from Robert Mann and Mitacs Globalink
Graduate Fellowship. J.Y. would like to thank Ahmadreza Moradpouri, Jianhui Qiu and Cameron Bunney for helpful discussions. We are grateful to the anonymous referees for the constructive feedback, which significantly enhanced the quality of this work. M. T. acknowledges the Canadian Institute for Theoretical Astrophysics (CITA) for financial support. This research is supported in part by the Natural Sciences and Engineering Research Council of Canada and the Perimeter Institute for Theoretical Physics. Research at Perimeter Institute is supported in part by the Government of Canada through the Department of Innovation, Science, and Economic Development and by the Province of Ontario through the Ministry of Colleges and Universities.

\bibliographystyle{JHEP}
\bibliography{biblio.bib}
\end{document}